%% file: main.tex
\documentclass[conference]{IEEEtran}
\IEEEoverridecommandlockouts

\input{preamble}

\begin{document}
\title{No physics required !\\
A visual-based introduction to GKP qubits for computer scientists}

\author{\IEEEauthorblockN{Richard A. Wolf}
\IEEEauthorblockA{
\textit{Irish Center for High-End Computing}\\
Galway, Ireland}
\and
\IEEEauthorblockN{Pavithran Iyer}
\IEEEauthorblockA{
\textit{Institute for Quantum Computing}\\
Waterloo, Canada}
}

\maketitle

\begin{abstract}
With the significance of continuous-variable quantum computing increasing thanks to the achievements of light-based quantum hardware, making it available to learner audiences outside physics has been an important yet seldom-tackled challenge. Similarly, the rising focus on fault-tolerant quantum computing has shed light on quantum error correction schemes, turning it into the locus of attention for industry and academia alike. In this paper, we explore the widely adopted framework of quantum error correction based on continuous variable systems and suggest a guide on building a self-contained learning session targeting the famous Gottesman-Kitaev-Preskill (GKP) code through its geometric intuition.
\end{abstract}


\input{sections/intro}

\input{sections/session_outline}
\input{sections/conclusion}

{\small
\bibliographystyle{abbrv}
\bibliography{main}
}

\end{document}

%% file: preamble.tex
\usepackage{cite}
\usepackage{amsmath,amssymb,amsfonts, txfonts}
\usepackage{algorithmic}
\usepackage{graphicx}
\usepackage{textcomp}
\usepackage{xcolor}
\usepackage{braket}
\usepackage{booktabs}
\usepackage{multirow}
\usepackage{pifont}
\usepackage{tablefootnote}
\usepackage{caption, subcaption}
\usepackage[export]{adjustbox}
\usepackage{listing}
\usepackage{tcolorbox}
\usepackage{graphicx}
\usepackage{tikz}
\usepackage{wrapfig}
\usepackage[ruled]{algorithm2e} 
\SetKwInput{KwInput}{Input}
\usepackage{bbold}

\def\BibTeX{{\rm B\kern-.05em{\sc i\kern-.025em b}\kern-.08em
    T\kern-.1667em\lower.7ex\hbox{E}\kern-.125emX}}

%% file: sections/intro.tex
\section{Introduction}

With quantum being considered the next computational revolution, a growing number of people with a background in classical computing have been broadening the scope of their expertise by delving into quantum computing \cite{Dimitrov:2023bjw}. For this audience seeking a bridge between their existing classical computer science (CS) expertise and the field of quantum computing and quantum information \cite{seegerer2021quantum, rieffel2000introduction}, the challenge isn't so much about building a skill-set from the ground up as it is about creating conceptual bridges allowing them to exploit their existing skills to their full extent in the pursuit of this novel field of computation.
In this regard, a number of quality material exists for tackling discrete-variable quantum computing with emphasis on fundamentals of linear algebra and probabilities and statistics, both well-known to classical computer scientists \cite{seegerer2021quantum}. However, the field of continuous-variable quantum computing (CVQC) remains relatively opaque and sub-exploited for learners with a classical CS background due to the simple fact that most if not all the introductory material to this type of computation is geared towards physicists. This poses a double problem. First and perhaps most intuitively, this poses a problem for motivated learners who lack a physics background. Instead of having the opportunity to harness their existing skills to tackle a new piece of knowledge, they are forced to build an understanding of this new computational paradigm of CVQC from the ground up, involving a major overhead in the acquisition of a number of physics-related notions. Second and perhaps most importantly, this is directly detrimental to the field of CVQC itself. Having a single point of access through physics results in the vast majority of people working in the field coming from  similar educational backgrounds. Where diversity of background is often a key element in the flourishing of innovation \cite{Alvargonzález01122011}, this homogenising effect of the single entry point could lead to the field missing out on valuable inputs it could receive from classical computer scientists \cite{doi:10.1177/00187208211048301}. It is thus not only in the interest of keen learners themselves but also in that of the field in itself \cite{martonosi2019stepsquantumcomputingcomputer} that we wish to foster an alternative way of approaching CVQC that does not rely on physics.

The learning material is guided by a series of visuals specifically designed to enhance intuitive understanding of the continuous variable (CV) and GKP encoding, and building on the classical coding theory skills learners from CS will come equipped with. While leveraging visual learning was a natural choice to convey the geometric intuitions behind GKP qubits, it also offers other advantages over the more traditional text-and-formula-based learning. By engaging visual communication brain regions, it has been shown to boost high-order thinking skills \cite{raiyn2016role} and promote higher research quality for those engaging with it \cite{mcgrath2005visual}. The visual approach has also displayed promising impact on engagement of under-served and under-represented minorities in STEM \cite{mcgrath2005visual}. While the interplay between visual content and quantum computing had attracted previous attention \cite{archer2022visual, Deveney:23, folkersvisualising}, with perhaps its most famous representative being the Bloch sphere, to the best of our knowledge no efforts had been made to harness those in the service of CVQC.

Through this paper, we seek to offer those designing content for learners with a background in CS with a self-contained introductory activity on CVQC specifically geared towards this audience and devoid of physics-based requirements. Prerequisites are limited to undergraduate-level linear algebra and probability theory together with basics of classical coding theory, all of which are a staple for undergraduate computer scientists. Additionally, we assume basic familiarity with complex numbers, which includes their representation as well as basic arithmetic operations.

In the following section, we suggest a gentle and CS-friendly pathway through the CVQC formalism, from information encoding to decoding and to the GKP qubit.

%% file: sections/session_outline.tex
\section{Session outline}
We present the outline of a two-hour session aimed at conveying the geometric intuition behind GKP qubits, making them an intuitive and natural solution to the problem of encoding continuous quantum information. 

\subsection{How can you encode quantum information?}
In this first section, the basics of quantum information are introduced in a manner aimed at facilitating later intuition for the GKP qubit.
Our goal is to introduce basic terminologies in quantum information that are essential to follow the rest of the material. This exposition is meant to be intuitive and assumes familiarity with classical coding theory concepts like encoding information, logical gates and errors.

\subsubsection{Learning objectives}
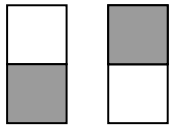
\begin{figure}[htbp]
    \centering
    \input{figures/2lvl_information}
        \caption{An example two-level encoding of two logical values. Here assuming a logical zero denoted $\ket{0_L}$ on the left and a logical one denoted $\ket{1_L}$ on the right.}
    \label{fig:2lvl-info}
\end{figure}

\begin{itemize}
    \item Qubits \begin{itemize}
        \item The notion of a qubit as having two states (Fig. \ref{fig:2lvl-info}), akin to a classical bit.
        \item Based on classical coding theory, the logical states $\ket{0_L}$ and $\ket{1_L}$, which are shown in the illustration in Fig. \ref{fig:2lvl-info}.
        \item The information stored in a qubit is formally written as a superposition of the two logical states $\ket{0_{L}}$ and $\ket{1_{L}}$, expressed as $\alpha\ket{0_{L}} + \beta \ket{1_{L}}$, where $\alpha$ and $\beta$ can be any complex numbers, such that $|\alpha|^2 + |\beta|^2 = 1$. An encoded state is characterized by a particular value of $\alpha$ and $\beta$.
    \end{itemize}
    \item $\mathbf{X}-$type errors are the quantum analogue of classical bit-flip errors. In the schematic level diagram (Fig. \ref{fig:2lvl-state-Xerr}) an $\mathbf{X}-$error corresponds to a displacement of the active cell by one.
    \item A logical $\mathbf{X}-$error, denoted by $\mathbf{X_L}$ acts on the encoded quantum information as as $\mathbf{X_L}:\ket{0_L} \mapsto \ket{1_L}$ and $\mathbf{X_L}:\ket{1_L} \mapsto \ket{0_L}$.
    \item The difference between a \emph{physical error} (denoted $\mathbf{X}$) and a \emph{logical error} (denoted $\mathbf{X}_L$) can be introduced here, drawing analogies from the classical coding theory. We only expect to be able to recover from physical errors and not from logical ones.
    \item In this encoding, a single displacement: $\mathbf{X}$ is also a logical error $\mathbf{X}_{L}$ (Fig. \ref{fig:2lvl-state-Xerr}). Hence the two-level encoding isn't robust against an $\mathbf{X}$ error.
\end{itemize}

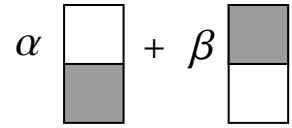
\begin{figure}[htbp]
    \centering
    \input{figures/2lvl_qubit}
        \caption{An example quantum state where a two-level encoding is used. Here the quantum logical zero denoted $\ket{0_L}$ has a probability amplitude of $\alpha$ while the quantum logical one denoted $\ket{1_L}$ has a probability amplitude of $\beta$.}
    \label{fig:2lvl-state}
\end{figure}

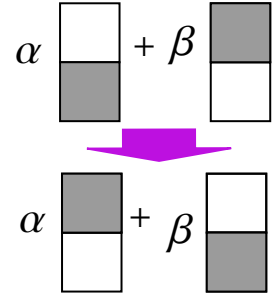
\begin{figure}[htbp]
    \centering
    \input{figures/2lvl_state_Xerr}
        \caption{The action of a single $\mathbf{X}$ error is illustrated on the example quantum state presented in Fig. \ref{fig:2lvl-state}. Here $\mathbf{X}(\ket{0_L})\mapsto\ket{1_L}$ and $\mathbf{X}(\ket{1_L})\mapsto\ket{0_L}$ thus causing a logical error.}
    \label{fig:2lvl-state-Xerr}
\end{figure}

\subsubsection{Estimated time}
12 minutes (12/120).

\subsubsection{Remarks and suggestions}
\begin{itemize}
    \item Chapter 1 of \cite{nilson2016teaching} provides a rigorous and detailed coverage of the concepts covered in this section.
    \item We will be using a level diagram similar to Fig. \ref{fig:2lvl-info} throughout to denote a quantum state. In this diagram, we indicate an \emph{active level} by a shaded cell, and refer to the corresponding quantum state as $\ket{i}$ where $i$ is the shaded cell. In this section, we use diagrams with only two levels denoting states as $\ket{0}$ and $\ket{1}$. However, when we have multiple levels, we can also use diagrams to represent $\ket{i}$ for $i \geq 1$.
    \item The concept of superposition is likely the most challenging and crucial at this point. A common misconception might be that both values (for 1 and 0) could be encoded separately, it is therefore important here to clarify that those variables are related to each other and storing them separately would entail losing important relations, if this remains still confusing it might be necessary to explain the change of basis.
\end{itemize}


\subsection{How can you make errors detectable?}
In this section, we aim to focus on the effect of errors in the encoded quantum information (qubit) and discuss ways of detecting the effect of an error. Here, learners can build on their background in classical coding theory to tackle the task of encoding information through redundancy to make it more resistant to errors.
The goal of this section is to introduce the basic terminology of quantum error correction.

\subsubsection{Learning objectives}
\begin{itemize}
    \item Redundancy, which can be familiar through the classical repetition code, is added in terms of levels of separation between the two levels encoding logical values. Adding \textit{padding} between levels is what improves the encoding.
    \item The terminology of \emph{encoding a qubit using $N$ levels} as the process of adding $N-2$ padding levels between the levels encoding the logical values.
    \item The concept of \emph{code space} as corresponding to the levels where the logical values $|0_{L}\rangle$ and $|1_{L}\rangle$ are encoded, and its complement where non-logical values fall.
    \item \emph{Detecting} errors, likely familiar from classical coding theory.
    \item Motivating why 3 levels is insufficient for recovery, since recovery is synonymous with inferring the (unknown) error that affected the encoded information.
    \item The equivalence between detecting the presence of an error and in this case asking the question \textit{``What is the active level $\mod~(N-1)$?"} for $N$ levels. 
    \item The \emph{syndrome} as the value of the active error $\mod~(N-1)$ for an $N$-level encoding.
\end{itemize}

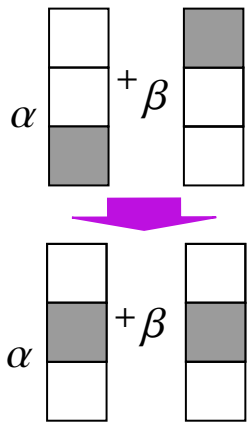
\begin{figure}[htbp]
    \centering
    \input{figures/3lvl_state_Xerr}
        \caption{The action of a single $\mathbf{X}$ error illustrated on an example quantum state where a three-level encoding is used. On the left side we have $\ket{0_L}$ with an active cell at the bottom while on the right side we have $\ket{1_L}$ with an active cell on the top. Here $\mathbf{X}(\ket{0_L})\mapsto\ket{u}$ and $\mathbf{X}(\ket{1_L})\mapsto\ket{u}$ where $\ket{u}$ stands for an undefined logical state. Note that here $\mathbf{X}(\ket{0_L}) = \mathbf{X}(\ket{1_L})$.}
    \label{fig:3lvl-state-Xerr}
\end{figure}

\begin{figure}[htbp]
    \centering
    \input{figures/codespace}
        \caption{For a three-level encoding where $\ket{0_L}$ would be represented by a coloured cell at the bottom and $\ket{1_L}$ by a coloured cell at the top, the \emph{code-space} corresponds to both levels encoding the logical information in green, while the complement of the code-space corresponds to the undefined logical state in red.}
    \label{fig:3lvl-codepsace}
\end{figure}
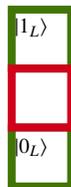

\subsubsection{Estimated time}
12 minutes (24/120).

\subsection{How could you decode those errors classically?}
Here we set the scene for the process of quantum decoding by harnessing one last time learner's existing mastery of classical coding theory. By extending the previous three-level encoding with a single level of \emph{padding} to a four-level one with now two levels of \emph{padding}, learners can build the intuition of the importance of \emph{padding} distance between levels encoding logical information.

\subsubsection{Learning objectives}
\begin{itemize}
    \item The difference between \emph{detection} -covered in the previous section- and possibility of \emph{correction} presented here. While correcting necessarily implies detecting, the reverse isn't true.
    \item Visually correcting $\mathbf{X}$ errors by bringing the shaded cell back to the closest extremity. The key idea here is to focus on visualising those shared blocks as slideable to correct information, as this will be necessary for the next sections.
    \item The concept of \emph{most likely} error, based on minimizing the number of $\mathbf{X}$ errors required to produce the observed pattern. In other words, we should focus on recovering from the most likely errors.
    \item Concept of \emph{decoding} as given a syndrome, identifying the most likely error that caused it. Parallels can be drawn with classical coding theory.
\end{itemize}

\begin{figure}[htbp]
    \centering
    \input{figures/4lvl_state_Xerr}
        \caption{The action of a single $\mathbf{X}$ error illustrated on an example quantum state where a four-level encoding is used. On the left side we have $\ket{0_L}$ with an active cell at the bottom while on the right side we have $\ket{1_L}$ with an active cell at the top. Here $\mathbf{X}(\ket{0_L})\mapsto\ket{u}$ and $\mathbf{X}(\ket{1_L})\mapsto\ket{u'}$ where $\ket{u}$ and $\ket{u'}$ stand for undefined logical states. The code-space is in green.}
    \label{fig:4lvl-state-Xerr}
\end{figure}
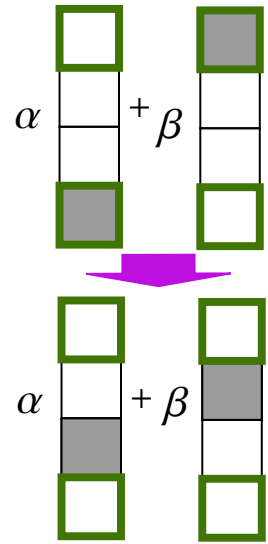

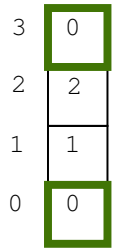
\begin{figure}[htbp]
    \centering
    \input{figures/4lvl_mod3}
        \caption{A four-level encoding illustrating a \textit{level modulo something} question style. The $k$ levels are listed on the left, while inside each level is the result of $ k \: mod \: 3$. Code-space in green.}
    \label{fig:4lvl-mod3}
\end{figure}

\subsubsection{Estimated time} 15 minutes (39/120).

\subsubsection{Remarks and suggestions}
\begin{enumerate}
    \item Emphasis on the idea that the only available information is the syndrome. The actual error is not accessible, only the syndrome is and the actual active level can never be revealed.
\end{enumerate}

\subsection{How can you decode in the context of quantum computing?}
This step represents the first conceptual jump away from classical coding theory and into quantum. The idea that \textit{some} measurements will destroy the information while others won't is crucial. The fact that measurements on the encoded state can potentially destroy the encoded information is an important quantum feature that has no classical counterpart and thus deserves careful attention.

Here learners will be guided in designing an encoding scheme based on the padding concept, setting the groundwork for the decoding algorithm Alg. \ref{algo:binning_decoder}.

\begin{algorithm}[h]
    \KwInput{$k \in \mathbb{N}^{+}$}
    $l \leftarrow$ location coordinate\\
    $a \leftarrow l \mod k $\\
    \If{$a \neq 0$}{
        $l \leftarrow l - a$\\
        \If{$a > \lceil\frac{k}{2}\rceil$}{
        $l \leftarrow l + k$
        }
    }
    \caption{Error correction algorithm - binning decoder}
    \label{algo:binning_decoder}
\end{algorithm}

\begin{figure}[htbp]
    \centering
    \input{figures/different_error_same_measurement}
        \caption{Different errors can result in the same observable error \emph{syndrome}. Each cell is marked by the syndrome it would yield, the codespace is bordered in green. Here, three different possible errors are shown by the grey cells, all yielding the same error syndrome of $2$.}
    \label{fig:different-errors-same-measurement-res}
\end{figure}
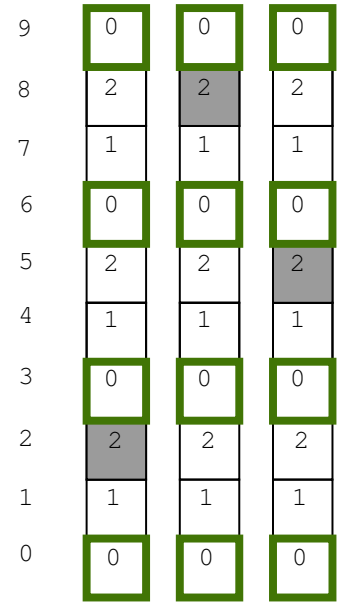

\subsubsection{Learning objectives}
\begin{itemize}
    \item Going back to the two-level system in Fig. \ref{fig:2lvl-state}, a measurement would correspond to the question: \textit{``Is the state $|0_L\rangle$ or $|1_L\rangle$?"}, which yields $|0_L\rangle$ with a probability $|\alpha|^{2}$ and $|1_L\rangle$ with a probability $|\beta|^{2}$.
    \item The \textit{stickiness} of measurement results: the idea that in quantum computing once a state is measured the result it has been \textit{observed to be in}  will stick forever is in stark contrast with the classical picture where accessing a classical binary sequence does not affect it. Circling back to the notion of superposition, the issue here is that the information contained in the superposition is lost forever since once the state is measured to be in either $\ket{0_L}$ or $\ket{1_L}$ it will stay in this forever. 
    
    \item The concept of \emph{syndrome measurement} as those measurements which do not distinguish between $\ket{0_L}$ and $\ket{1_L}$. Building a \textit{good question} corresponds to building a function $m(x) $ such that $m(\ket{0_L}) = m(\ket{1_L}) \neq m(x~\forall x \not\in \{0,1\})$. The necessity to pick the correct $mod$ for the syndrome measurement is directly related to the concept of code space and of the stickiness of measurement outcomes. Choosing the correct $mod$ value means that though the measurement outcome will be sticky, it does not affect the actual logical information because it still leaves the superposition intact within the code space. This is the reason behind the absolute necessity to avoid distinguishing between $\ket{0_L}$ and $\ket{1_L}$. Due to the way syndrome measurement are built, several different errors can correspond to the same outcome. The measurement function $m(x)$ is surjective but not injective, see Fig. \ref{fig:different-errors-same-measurement-res}. 
\end{itemize}

\subsubsection{Estimated time} 18 minutes (57/120)

\subsubsection{Remarks and suggestions}
\begin{itemize}
    \item The only set of \textit{allowed} measurements for a state $|\psi = \alpha|0\rangle + \beta|1\rangle$ are of the form \textit{``is the state equal to $|0\rangle$ or $|1\rangle$ ?"}. For advanced learners, it can be pointed out that $|0\rangle$ and $|1\rangle$ can be replaced by any orthogonal states.
    
    \item In the quantum setting, information must remain hidden throughout the detection and recovery process. It is important for learners to grasp this crucial distinction between quantum and classical error correction. A useful way to build intuition is to imagine adding a constraint to the classical recovery problem. Treating the detection and recovery operations as eavesdroppers. In this picture, the encoded information is private and must remain inaccessible to the recovery procedure.
    \item Suggestions to convey the intuition of the algorithm. A possible way is to frame the problem as the decoder would face it. Taking an example of $5$ levels and asking the question: \textit{``Given that the active level $\mod 4 = 2$, determine the error that caused this measurement of this syndrome?"} We could iterate on increasing number of levels to drive the intuition home or leave it at a single example.
    \item Logical errors occur if there are more shifts than half the padding space. Because the first step of decoding would apply an unintended logical operation.
\end{itemize}

\subsection{How to protect quantum information from the second type of errors?}
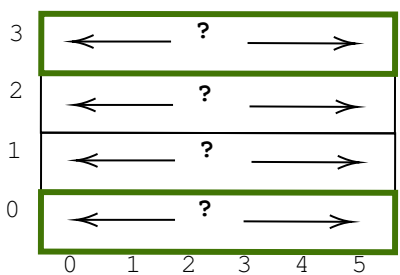
\begin{figure}[htbp]
    \centering
    \input{figures/4lvl_2axis_no_info}
        \caption{The uncertainty principle dictates that exact knowledge about the location in one axis (here vertical) results in infinite uncertainty about the other axis (here horizontal).}
    \label{fig:4lvl-2axis-noInfo}
\end{figure}

\begin{figure}[htbp]
    \centering
    \input{figures/10lvl_3multiples}
        \caption{An example encoding where $\ket{0_L}$ corresponds to even multiples of $3$, i.e. $\ket{0}, \ket{6}$ while $\ket{1_L}$ correspond to odd multiples of 3 i.e. $\ket{3}, \ket{9}$.}
    \label{fig:10lvl-3multiples}
\end{figure}
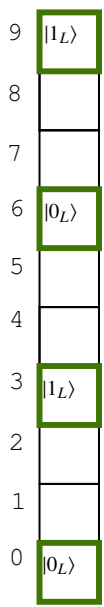

\begin{figure}[htbp]
    \centering
    \input{figures/error_transitions_diagram}
        \caption{There are 2 types of errors that we need to be able to detect and correct: $\mathbf{X}$ and $\mathbf{Z}$. The following state-transition diagram shows the action of each error on both $\ket{0_L}$ and $\ket{1_L}$}
    \label{fig:4lvl-X-and-Z-errors}
\end{figure}
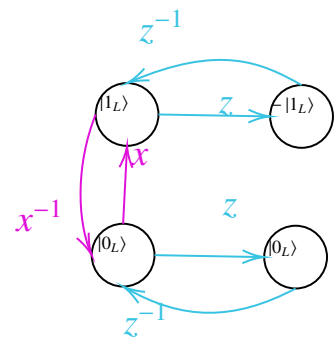

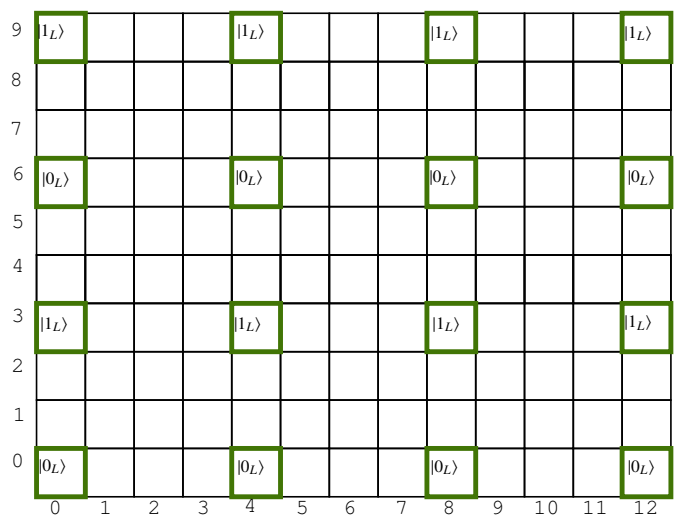
\begin{figure}[htbp]
    \centering
    \input{figures/10x13lvl_full_grid}
        \caption{Illustration that the padding doesn't need to be the same along both axis. Here for instance, logical information is encoded with $2$ levels of padding along the vertical axis but with $3$ along the horizontal axis. The code-space is in green. This entails that here the code-space is preserved by asking about the location modulo $3$ for the vertical axis where $\mathbf{X}$ errors happen and about the location modulo $4$ for the horizontal axis where $\mathbf{Z}$ errors happen.}
    \label{fig:10lx13lvl}
\end{figure}

\subsubsection{Learning objectives}
\begin{itemize}
    \item Quantum information, in contrast to its classical counterpart, experiences two different kinds of error: $\mathbf{X}$ and $\mathbf{Z}$ errors (Fig. \ref{fig:4lvl-X-and-Z-errors}).
    \item Extending the linear level diagram into a 2D grid, where $\mathbf{X}$ errors displace quantum information in the vertical axis while $\mathbf{Z}$ errors displace it in the horizontal axis (Fig. \ref{fig:10lx13lvl}).
    \item $\mathbf{Z}$ errors transform $|1_L\rangle$ to $-|1_L\rangle$ hence, a separation between $\ket{1_L}$ and $-\ket{1_L}$ on the horizontal axis is necessary to protect against $\mathbf{Z}$ errors.
\end{itemize}

\subsubsection{Estimated time} 16 minutes (73/120)

\subsubsection{Remarks and suggestions}
\begin{itemize}
\item The fact that $\mathbf{Z}$ errors seem to only affect $\ket{1_L}$ could be difficult for learners to grasp. If the transition diagram in Fig. \ref{fig:4lvl-X-and-Z-errors} isn't enough, it might be necessary to introduce the change of basis to make $\mathbf{Z}$ error effects more visible.
\end{itemize}

\subsection{How to respect the uncertainty principle?}
This section introduces an encoding scheme where  $|0\rangle$ and $|1\rangle$ correspond to a superposition of different cells. It represents a transition from the discrete level system to the CV setting in the following section.

\subsubsection{Learning objectives}
\begin{itemize}
    \item The vertical and horizontal location of the cell corresponding to $|0_L\rangle$ cannot be accurately specified. This is the limitation introduced by the uncertainty principle (Fig. \ref{fig:4lvl-2axis-noInfo}). In other words, if we fixed the vertical location of a level to denote $|0_L\rangle$, then the uncertainty about the horizontal location of the corresponding cell would be infinite. Consequently, there would be no protection against $\mathbf{Z}$ errors.
    \item To respect the uncertainty principle and have protection against both $\mathbf{X}$ and $\mathbf{Z}$ errors, we associate more than one level to $|0_L\rangle$ and more than one level to $|1_L\rangle$, both in the vertical and horizontal axes (Fig. \ref{fig:10lx13lvl}).
    \item In Fig. \ref{fig:10lx13lvl}, the logical $|0_L\rangle$ and $|1_L\rangle$ do not correspond to specific cells on any of the vertical or horizontal axes. Instead, logical $|0_L\rangle$ and $|1_L\rangle$ are equally distributed across multiple levels. Formally, we note that $|0_L\rangle$ is a superposition of levels $0$ and $6$ and likewise  $|1_L\rangle$ is a superposition of levels $3$ and $9$, the mathematical expression would be:
    \begin{gather}
        \ket{0_{L}} = \dfrac{|0\rangle + |6\rangle }{\sqrt{2}} \\
        \ket{1_{L}} = \dfrac{|3\rangle + |9\rangle}{\sqrt{2}} \label{eq:shift_resistant_code}
    \end{gather}
    \item Logical operations $\mathbf{X}_L$ and $\mathbf{Z}_L$ in Fig. \ref{fig:10lx13lvl} correspond to the vertical shifts by $3$: $\mathbf{X}^3$ and the horizontal shifts by $4$: $\mathbf{Z}^4$, respectively.
    \item Horizontal and vertical spacing does not need to be identical. This can give different levels of robustness against $\mathbf{X}$ and $\mathbf{Z}$ errors. Hence, $\mathbf{X}_L$ and $\mathbf{Z}_L$ errors correspond to different amounts of shifts (Fig. \ref{fig:10lx13lvl}). There is however a constraint which must be satisfied between the spacing on the horizontal and vertical axes, namely that the logical operators must anti-commute: $\{\mathbf{X}_L, \mathbf{Z}_L\} = 0$.
    \item A way to encode a qubit in levels $n$: choosing $k < n$, considering the superpositions of an even multiple of $k$ to denote $|0_L\rangle$, and a superposition of the odd multiples of $k$ to denote $|1_L\rangle$.
    \item The decoding algorithm presented in Alg \ref{algo:binning_decoder} to correct for $\mathbf{X}$ errors can also be used in a similar way to decode for $\mathbf{Z}$ errors. Decoding can be done for both error types independently along each axis to correct for its respective error.
\end{itemize}

\subsubsection{Estimated time} 22 minutes (95/120)

\subsubsection{Remarks and suggestions}
\begin{itemize}
\item The code developed in \ref{eq:shift_resistant_code} is called a \emph{shift-resistant code}, since it protects against vertical and horizontal shifts, i.e., $\mathbf{X}$ and $\mathbf{Z}$ errors.
\end{itemize}

\subsection{How can we generalize the encoding strategy to infinite levels?}
In this section we consider a CV encoding scheme for a qubit, where the level diagram is replaced by a continuous real line, and displacements are parametrized by a continuous variable. This is the last and most central section which finally introduces the famous Gottesman Kitaev Preskill (GKP) code.

\subsubsection{Learning objectives} 

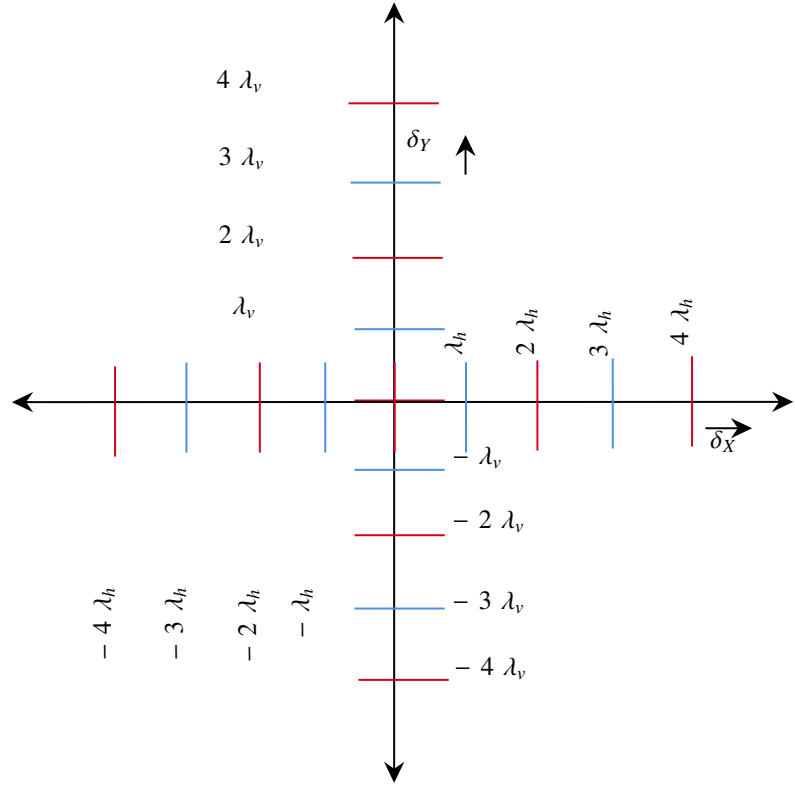
\begin{figure}
\centering
\input{figures/continuous_variables_XY.tex}
\caption{Figure showing the continuous variable generalization to the encoding scheme that corrects against small displacements in the 2D plane. The logical states are defined as infinite superpositions of states corresponding to values on the horizontal axis separated by a fixed spacing $\lambda_{h}$ and values on the vertical axis separated by $\lambda_{v}$.}
\label{fig:continuous_variables_XY}
\end{figure}

\begin{itemize}

\item Revisiting the one-dimensional levels in Fig. \ref{fig:10lvl-3multiples}, they can be extended to the CV setting using a twofold method.
\begin{itemize}
    \item First, by extending the discrete encoding to infinite levels, where $|0\rangle$ is encoded as an infinite superposition of all levels separated by 6, i.e.,
    \begin{flalign}
        \ket{0_L} &= \ldots + |-12\rangle + |-6\rangle + |0\rangle + |6\rangle + |12\rangle + \ldots  \label{eq:infinite_levels_discrete_0L} \\
        \ket{1_L} &= \ldots + |-15\rangle + |-9\rangle + |-3\rangle + |3\rangle + |9\rangle + \ldots \label{eq:infinite_levels_discrete_1L}
    \end{flalign}
    \item Second, taking the limit of the separation between two cells to zero, that is, a continuous real line wherein the $\mathbf{X}$ errors parametrized by $\delta_{v}$ cause a displacement by $\delta_{v}$ on the vertical axis. We will represent this as $\mathbf{X}(\delta_{v})$.
\end{itemize}

\item To protect against small shifts in the vertical axis, the logical states of the encoded qubit correspond to superpositions of points on  the vertical axis, that are separated by an amount $\lambda_v$:
    \begin{flalign}
        \ket{0_L} &= \sum_{n = -\infty}^{\infty}\ket{2n ~ \lambda_v} \label{eq:GKP_0L_vertical} \\
        \ket{1_L} &= \sum_{n = -\infty}^{\infty}\ket{(2n+1) ~ \lambda_v} \label{eq:GKP_1L_vertical}
    \end{flalign}
    In other words, we have infinite superpositions over the odd and even multiples of $\lambda_v$ to constitute $\ket{0_L}$ and $\ket{1_L}$ respectively.

\item Any displacement along the vertical axis $\delta_{v}$ can be corrected as long as $\delta_{v} < \frac{\lambda_v}{2}$. Similarly any displacement along the horizontal axis $\delta_{h}$ can be corrected as long as $\delta_{h} < \frac{\lambda_h}{2}$.

\item $\mathbf{X}_L$ that transforms $|0_L\rangle$ to $|1_L\rangle$ is a displacement by a fixed amount: $\mathbf{X}_L = \mathbf{X}(\lambda_v)$ (Fig. \ref{fig:continuous_variables_XY}). Similarly $\mathbf{Z}_L$ is a displacement by a fixed amount: $\mathbf{Z}_L = \mathbf{Z}(\lambda_h)$ (Fig. \ref{fig:continuous_variables_XY}).

\item The spacings $\lambda_{v}$ and $\lambda_{h}$ determine how robust the encoding scheme is to displacement errors of the $\mathbf{X}$ and $\mathbf{Z}$ type.

\item The horizontal and the vertical spacings: $\lambda_{h}$ and $\lambda_{v}$, cannot be arbitrarily set. They are related by the uncertainty principle. If $\lambda_{h}$ grows larger, $\lambda_{v}$ must be smaller.

\item The anti-commutation relation $\{\mathbf{X}, \mathbf{Z}\} = 0$ implies that $\lambda_{h} \cdot \lambda_{v} = \pi$ \cite{GKP01,GP21}. 

\item A popular solution for both values of $\lambda_v$ and $\lambda_h$ is $\lambda_{h} = \lambda_{v} = \sqrt{\pi}$, the \emph{Gottesman Kitaev Preskil (GKP)} code \cite{GKP01}.

\end{itemize}

\begin{figure}
\begin{center}
\input{figures/continuous_X_level.tex}
\caption{Figure showing the infinite level diagram that generalizes the encoding scheme discussed so far to continuous variables. $\mathbf{X}$ errors are depicted using displacements on the vertical axis, parametrized by a real number $\delta_{v}$. Logical operations are fixed displacements, i.e., $\mathbf{X}(\lambda_{v})$.}
\label{fig:continuous_X_level}
\end{center}
\end{figure}
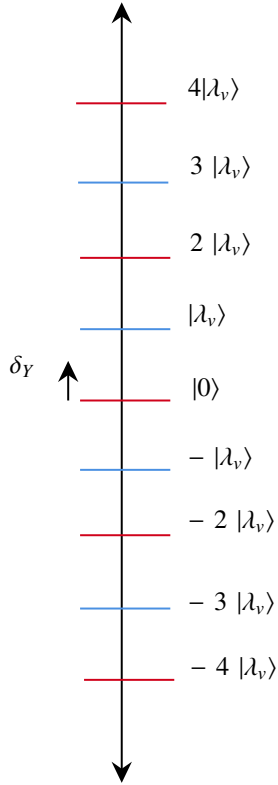

\subsubsection{Estimated time}  25 minutes (120/120)

\subsubsection{Remarks and suggestions}
\begin{itemize}
    \item Beyond the popular GKP, a more general solution which includes different capabilities for correcting $\mathbf{X}$ and $\mathbf{Z}$ errors is discussed in \cite{A22,GP21}.

    \item The question of the normalisation of GKP states is subtle and falls outside the scope of this introductory session setup. However, for comfortable learners, it will be important to note that the states in equations \ref{eq:infinite_levels_discrete_0L} and \ref{eq:infinite_levels_discrete_1L} are not normalized since they are infinite superpositions. A detailed discussion of approximate GKP states can be found in \cite{GKP01,MYHK20,WBZ23}.

\end{itemize}

%% file: figures/2lvl_information.tex
\tikzset{every picture/.style={line width=0.75pt}} 

\begin{tikzpicture}[x=0.75pt,y=0.75pt,yscale=-1,xscale=1]

\draw   (20,10.25) -- (50,10.25) -- (50,40) -- (20,40) -- cycle ;
\draw  [fill={rgb, 255:red, 155; green, 155; blue, 155 }  ,fill opacity=1 ] (20,40) -- (50,40) -- (50,69.75) -- (20,69.75) -- cycle ;
\draw   (71,10.25) -- (101,10.25) -- (101,40) -- (71,40) -- cycle ;
\draw   (71,40) -- (101,40) -- (101,69.75) -- (71,69.75) -- cycle ;
\draw  [fill={rgb, 255:red, 155; green, 155; blue, 155 }  ,fill opacity=1 ] (71,10.25) -- (101,10.25) -- (101,40) -- (71,40) -- cycle ;

\end{tikzpicture}

%% file: figures/2lvl_qubit.tex
\tikzset{every picture/.style={line width=0.75pt}} 

\begin{tikzpicture}[x=0.75pt,y=0.75pt,yscale=-1,xscale=1]

\draw   (277,9.25) -- (307,9.25) -- (307,39) -- (277,39) -- cycle ;
\draw  [fill={rgb, 255:red, 155; green, 155; blue, 155 }  ,fill opacity=1 ] (277,39) -- (307,39) -- (307,68.75) -- (277,68.75) -- cycle ;
\draw   (360,9.25) -- (390,9.25) -- (390,39) -- (360,39) -- cycle ;
\draw   (360,39) -- (390,39) -- (390,68.75) -- (360,68.75) -- cycle ;
\draw  [fill={rgb, 255:red, 155; green, 155; blue, 155 }  ,fill opacity=1 ] (360,9.25) -- (390,9.25) -- (390,39) -- (360,39) -- cycle ;

\draw (251,23) node [anchor=north west][inner sep=0.75pt]  [font=\LARGE] [align=left] {$\displaystyle \alpha $};
\draw (340,20) node [anchor=north west][inner sep=0.75pt]  [font=\LARGE] [align=left] {$\displaystyle \beta $};
\draw (315,25) node [anchor=north west][inner sep=0.75pt]   [align=left] {{\Large +}};

\end{tikzpicture}

%% file: figures/2lvl_state_Xerr.tex
\tikzset{every picture/.style={line width=0.75pt}} 

\begin{tikzpicture}[x=0.75pt,y=0.75pt,yscale=-1,xscale=1]

\draw   (496,379.25) -- (526,379.25) -- (526,409) -- (496,409) -- cycle ;
\draw  [fill={rgb, 255:red, 155; green, 155; blue, 155 }  ,fill opacity=1 ] (496,409) -- (526,409) -- (526,438.75) -- (496,438.75) -- cycle ;
\draw   (572,379.25) -- (602,379.25) -- (602,409) -- (572,409) -- cycle ;
\draw   (572,409) -- (602,409) -- (602,438.75) -- (572,438.75) -- cycle ;
\draw  [fill={rgb, 255:red, 155; green, 155; blue, 155 }  ,fill opacity=1 ] (572,379.25) -- (602,379.25) -- (602,409) -- (572,409) -- cycle ;
\draw   (497,465.25) -- (527,465.25) -- (527,495) -- (497,495) -- cycle ;
\draw   (497,495) -- (527,495) -- (527,524.75) -- (497,524.75) -- cycle ;
\draw   (570,465.25) -- (600,465.25) -- (600,495) -- (570,495) -- cycle ;
\draw   (570,495) -- (600,495) -- (600,524.75) -- (570,524.75) -- cycle ;
\draw   (570,465.25) -- (600,465.25) -- (600,495) -- (570,495) -- cycle ;
\draw  [fill={rgb, 255:red, 155; green, 155; blue, 155 }  ,fill opacity=1 ] (497,465.25) -- (527,465.25) -- (527,495) -- (497,495) -- cycle ;
\draw  [fill={rgb, 255:red, 155; green, 155; blue, 155 }  ,fill opacity=1 ] (570,495) -- (600,495) -- (600,524.75) -- (570,524.75) -- cycle ;
\draw  [color={rgb, 255:red, 189; green, 16; blue, 224 }  ,draw opacity=1 ][fill={rgb, 255:red, 189; green, 16; blue, 224 }  ,fill opacity=1 ] (564.08,442.92) -- (564.11,452.42) -- (582.3,452.37) -- (545.94,458.81) -- (509.55,452.58) -- (527.73,452.53) -- (527.71,443.03) -- cycle ;

\draw (531,395) node [anchor=north west][inner sep=0.75pt]   [align=left] {{\Large +}};
\draw (529,481) node [anchor=north west][inner sep=0.75pt]   [align=left] {{\Large +}};
\draw (472,398) node [anchor=north west][inner sep=0.75pt]  [font=\LARGE] [align=left] {$\displaystyle \alpha $};
\draw (551,389) node [anchor=north west][inner sep=0.75pt]  [font=\LARGE] [align=left] {$\displaystyle \beta $};
\draw (550,484) node [anchor=north west][inner sep=0.75pt]  [font=\LARGE] [align=left] {$\displaystyle \beta $};
\draw (474,483) node [anchor=north west][inner sep=0.75pt]  [font=\LARGE] [align=left] {$\displaystyle \alpha $};

\end{tikzpicture}

%% file: figures/3lvl_state_Xerr.tex
\tikzset{every picture/.style={line width=0.75pt}} 

\begin{tikzpicture}[x=0.75pt,y=0.75pt,yscale=-1,xscale=1]

\draw   (337,1439.25) -- (367,1439.25) -- (367,1469) -- (337,1469) -- cycle ;
\draw  [fill={rgb, 255:red, 155; green, 155; blue, 155 }  ,fill opacity=1 ] (337,1469) -- (367,1469) -- (367,1498.75) -- (337,1498.75) -- cycle ;
\draw   (405,1439.25) -- (435,1439.25) -- (435,1469) -- (405,1469) -- cycle ;
\draw   (405,1469) -- (435,1469) -- (435,1498.75) -- (405,1498.75) -- cycle ;
\draw   (405,1439.25) -- (435,1439.25) -- (435,1469) -- (405,1469) -- cycle ;
\draw   (336,1528.25) -- (366,1528.25) -- (366,1558) -- (336,1558) -- cycle ;
\draw   (336,1558) -- (366,1558) -- (366,1587.75) -- (336,1587.75) -- cycle ;
\draw   (406,1528.25) -- (436,1528.25) -- (436,1558) -- (406,1558) -- cycle ;
\draw   (406,1558) -- (436,1558) -- (436,1587.75) -- (406,1587.75) -- cycle ;
\draw   (406,1528.25) -- (436,1528.25) -- (436,1558) -- (406,1558) -- cycle ;
\draw   (336,1528.25) -- (366,1528.25) -- (366,1558) -- (336,1558) -- cycle ;
\draw  [fill={rgb, 255:red, 155; green, 155; blue, 155 }  ,fill opacity=1 ] (406,1558) -- (436,1558) -- (436,1587.75) -- (406,1587.75) -- cycle ;
\draw   (336,1587.75) -- (366,1587.75) -- (366,1617.5) -- (336,1617.5) -- cycle ;
\draw   (406,1587.75) -- (436,1587.75) -- (436,1617.5) -- (406,1617.5) -- cycle ;
\draw  [fill={rgb, 255:red, 155; green, 155; blue, 155 }  ,fill opacity=1 ] (405,1409.5) -- (435,1409.5) -- (435,1439.25) -- (405,1439.25) -- cycle ;
\draw   (337,1409.5) -- (367,1409.5) -- (367,1439.25) -- (337,1439.25) -- cycle ;
\draw  [fill={rgb, 255:red, 155; green, 155; blue, 155 }  ,fill opacity=1 ] (336,1558) -- (366,1558) -- (366,1587.75) -- (336,1587.75) -- cycle ;
\draw  [color={rgb, 255:red, 189; green, 16; blue, 224 }  ,draw opacity=1 ][fill={rgb, 255:red, 189; green, 16; blue, 224 }  ,fill opacity=1 ] (403.08,1505.32) -- (403.11,1514.82) -- (421.3,1514.77) -- (384.94,1521.21) -- (348.55,1514.98) -- (366.73,1514.93) -- (366.71,1505.43) -- cycle ;

\draw (369,1439) node [anchor=north west][inner sep=0.75pt]   [align=left] {{\Large +}};
\draw (368,1559) node [anchor=north west][inner sep=0.75pt]   [align=left] {{\Large +}};
\draw (316,1457.67) node [anchor=north west][inner sep=0.75pt]  [font=\LARGE] [align=left] {$\displaystyle \alpha $};
\draw (384,1444) node [anchor=north west][inner sep=0.75pt]  [font=\LARGE] [align=left] {$\displaystyle \beta $};
\draw (383,1559) node [anchor=north west][inner sep=0.75pt]  [font=\LARGE] [align=left] {$\displaystyle \beta $};
\draw (314,1578.67) node [anchor=north west][inner sep=0.75pt]  [font=\LARGE] [align=left] {$\displaystyle \alpha $};

\end{tikzpicture}

%% file: figures/codespace.tex
\tikzset{every picture/.style={line width=0.75pt}} 

\begin{tikzpicture}[x=0.75pt,y=0.75pt,yscale=-1,xscale=1]

\draw   (523,2707.46) -- (553,2707.46) -- (553,2737.21) -- (523,2737.21) -- cycle ;
\draw   (523,2737.21) -- (553,2737.21) -- (553,2766.96) -- (523,2766.96) -- cycle ;
\draw   (523,2677.71) -- (553,2677.71) -- (553,2707.46) -- (523,2707.46) -- cycle ;
\draw  [color={rgb, 255:red, 65; green, 117; blue, 5 }  ,draw opacity=1 ][line width=3]  (523,2677.71) -- (553,2677.71) -- (553,2707.46) -- (523,2707.46) -- cycle ;
\draw  [color={rgb, 255:red, 65; green, 117; blue, 5 }  ,draw opacity=1 ][line width=3]  (523,2737.21) -- (553,2737.21) -- (553,2766.96) -- (523,2766.96) -- cycle ;
\draw  [color={rgb, 255:red, 208; green, 2; blue, 27 }  ,draw opacity=1 ][line width=3]  (523,2707.46) -- (553,2707.46) -- (553,2737.21) -- (523,2737.21) -- cycle ;

\draw (523,2680.71) node [anchor=north west][inner sep=0.75pt]  [font=\footnotesize] [align=left] {$\displaystyle \ket{1_{L}}$};
\draw (523,2741.21) node [anchor=north west][inner sep=0.75pt]  [font=\footnotesize] [align=left] {$\displaystyle \ket{0_{L}}$};

\end{tikzpicture}

%% file: figures/4lvl_state_Xerr.tex
\tikzset{every picture/.style={line width=0.75pt}} 

\begin{tikzpicture}[x=0.75pt,y=0.75pt,yscale=-1,xscale=1]

\draw   (358,5281.75) -- (388,5281.75) -- (388,5311.5) -- (358,5311.5) -- cycle ;
\draw   (358,5311.5) -- (388,5311.5) -- (388,5341.25) -- (358,5341.25) -- cycle ;
\draw   (429,5252.75) -- (459,5252.75) -- (459,5282.5) -- (429,5282.5) -- cycle ;
\draw   (429,5282.5) -- (459,5282.5) -- (459,5312.25) -- (429,5312.25) -- cycle ;
\draw  [color={rgb, 255:red, 65; green, 117; blue, 5 }  ,draw opacity=1 ][fill={rgb, 255:red, 155; green, 155; blue, 155 }  ,fill opacity=1 ][line width=3]  (429,5252.75) -- (459,5252.75) -- (459,5282.5) -- (429,5282.5) -- cycle ;
\draw   (429,5312.25) -- (459,5312.25) -- (459,5342) -- (429,5342) -- cycle ;
\draw  [color={rgb, 255:red, 65; green, 117; blue, 5 }  ,draw opacity=1 ][line width=3]  (358,5252) -- (388,5252) -- (388,5281.75) -- (358,5281.75) -- cycle ;
\draw  [color={rgb, 255:red, 65; green, 117; blue, 5 }  ,draw opacity=1 ][fill={rgb, 255:red, 155; green, 155; blue, 155 }  ,fill opacity=1 ][line width=3]  (358,5341.25) -- (388,5341.25) -- (388,5371) -- (358,5371) -- cycle ;
\draw  [color={rgb, 255:red, 65; green, 117; blue, 5 }  ,draw opacity=1 ][line width=3]  (429,5342) -- (459,5342) -- (459,5371.75) -- (429,5371.75) -- cycle ;
\draw  [color={rgb, 255:red, 189; green, 16; blue, 224 }  ,draw opacity=1 ][fill={rgb, 255:red, 189; green, 16; blue, 224 }  ,fill opacity=1 ] (426.08,5376.05) -- (426.11,5385.55) -- (444.3,5385.5) -- (407.94,5391.94) -- (371.55,5385.72) -- (389.73,5385.66) -- (389.71,5376.16) -- cycle ;
\draw   (359,5428.92) -- (389,5428.92) -- (389,5458.67) -- (359,5458.67) -- cycle ;
\draw   (359,5458.67) -- (389,5458.67) -- (389,5488.42) -- (359,5488.42) -- cycle ;
\draw   (430,5399.92) -- (460,5399.92) -- (460,5429.67) -- (430,5429.67) -- cycle ;
\draw   (430,5429.67) -- (460,5429.67) -- (460,5459.42) -- (430,5459.42) -- cycle ;
\draw   (430,5459.42) -- (460,5459.42) -- (460,5489.17) -- (430,5489.17) -- cycle ;
\draw  [color={rgb, 255:red, 65; green, 117; blue, 5 }  ,draw opacity=1 ][line width=3]  (359,5399.17) -- (389,5399.17) -- (389,5428.92) -- (359,5428.92) -- cycle ;
\draw  [color={rgb, 255:red, 65; green, 117; blue, 5 }  ,draw opacity=1 ][line width=3]  (430,5489.17) -- (460,5489.17) -- (460,5518.92) -- (430,5518.92) -- cycle ;
\draw  [fill={rgb, 255:red, 155; green, 155; blue, 155 }  ,fill opacity=1 ] (359,5458.67) -- (389,5458.67) -- (389,5488.42) -- (359,5488.42) -- cycle ;
\draw  [fill={rgb, 255:red, 155; green, 155; blue, 155 }  ,fill opacity=1 ] (430,5429.67) -- (460,5429.67) -- (460,5459.42) -- (430,5459.42) -- cycle ;
\draw  [color={rgb, 255:red, 65; green, 117; blue, 5 }  ,draw opacity=1 ][line width=3]  (359,5488.42) -- (389,5488.42) -- (389,5518.17) -- (359,5518.17) -- cycle ;
\draw  [color={rgb, 255:red, 65; green, 117; blue, 5 }  ,draw opacity=1 ][line width=3]  (430,5399.92) -- (460,5399.92) -- (460,5429.67) -- (430,5429.67) -- cycle ;

\draw (391,5295.5) node [anchor=north west][inner sep=0.75pt]   [align=left] {{\Large +}};
\draw (392,5442.67) node [anchor=north west][inner sep=0.75pt]   [align=left] {{\Large +}};
\draw (334,5300.67) node [anchor=north west][inner sep=0.75pt]  [font=\LARGE] [align=left] {$\displaystyle \alpha $};
\draw (407,5301) node [anchor=north west][inner sep=0.75pt]  [font=\LARGE] [align=left] {$\displaystyle \beta $};
\draw (335,5444.67) node [anchor=north west][inner sep=0.75pt]  [font=\LARGE] [align=left] {$\displaystyle \alpha $};
\draw (410,5441) node [anchor=north west][inner sep=0.75pt]  [font=\LARGE] [align=left] {$\displaystyle \beta $};

\end{tikzpicture}

%% file: figures/4lvl_mod3.tex
\tikzset{every picture/.style={line width=0.75pt}} 

\begin{tikzpicture}[x=0.75pt,y=0.75pt,yscale=-1,xscale=1]

\draw   (138,2452.08) -- (168,2452.08) -- (168,2481.83) -- (138,2481.83) -- cycle ;
\draw   (138,2481.83) -- (168,2481.83) -- (168,2511.58) -- (138,2511.58) -- cycle ;
\draw  [color={rgb, 255:red, 65; green, 117; blue, 5 }  ,draw opacity=1 ][line width=3]  (138,2422.33) -- (168,2422.33) -- (168,2452.08) -- (138,2452.08) -- cycle ;
\draw  [color={rgb, 255:red, 65; green, 117; blue, 5 }  ,draw opacity=1 ][line width=3]  (138,2511.58) -- (168,2511.58) -- (168,2541.33) -- (138,2541.33) -- cycle ;

\draw (116,2515.14) node [anchor=north west][inner sep=0.75pt]   [align=left] {{\fontfamily{pcr}\selectfont 0}};
\draw (117,2485.14) node [anchor=north west][inner sep=0.75pt]   [align=left] {{\fontfamily{pcr}\selectfont 1}};
\draw (118,2455.14) node [anchor=north west][inner sep=0.75pt]   [align=left] {{\fontfamily{pcr}\selectfont 2}};
\draw (118,2426.14) node [anchor=north west][inner sep=0.75pt]   [align=left] {{\fontfamily{pcr}\selectfont 3}};
\draw (145,2515.14) node [anchor=north west][inner sep=0.75pt]   [align=left] {{\fontfamily{pcr}\selectfont 0}};
\draw (145,2426.14) node [anchor=north west][inner sep=0.75pt]   [align=left] {{\fontfamily{pcr}\selectfont 0}};
\draw (145,2485.14) node [anchor=north west][inner sep=0.75pt]   [align=left] {{\fontfamily{pcr}\selectfont 1}};
\draw (146,2456.14) node [anchor=north west][inner sep=0.75pt]   [align=left] {{\fontfamily{pcr}\selectfont 2}};

\end{tikzpicture}

%% file: figures/different_error_same_measurement.tex
\tikzset{every picture/.style={line width=0.75pt}} 

\begin{tikzpicture}[x=0.75pt,y=0.75pt,yscale=-1,xscale=1]

\draw   (429,4227.93) -- (459,4227.93) -- (459,4257.68) -- (429,4257.68) -- cycle ;
\draw   (429,4257.68) -- (459,4257.68) -- (459,4287.43) -- (429,4287.43) -- cycle ;
\draw   (429,4227.93) -- (459,4227.93) -- (459,4257.68) -- (429,4257.68) -- cycle ;
\draw   (429,4317.18) -- (459,4317.18) -- (459,4346.93) -- (429,4346.93) -- cycle ;
\draw   (429,4346.93) -- (459,4346.93) -- (459,4376.68) -- (429,4376.68) -- cycle ;
\draw   (429,4346.93) -- (459,4346.93) -- (459,4376.68) -- (429,4376.68) -- cycle ;
\draw   (429,4406.43) -- (459,4406.43) -- (459,4436.18) -- (429,4436.18) -- cycle ;
\draw   (429,4436.18) -- (459,4436.18) -- (459,4465.93) -- (429,4465.93) -- cycle ;
\draw  [color={rgb, 255:red, 65; green, 117; blue, 5 }  ,draw opacity=1 ][line width=3]  (429,4198.18) -- (459,4198.18) -- (459,4227.93) -- (429,4227.93) -- cycle ;
\draw  [color={rgb, 255:red, 65; green, 117; blue, 5 }  ,draw opacity=1 ][line width=3]  (429,4465.93) -- (459,4465.93) -- (459,4495.68) -- (429,4495.68) -- cycle ;
\draw  [color={rgb, 255:red, 65; green, 117; blue, 5 }  ,draw opacity=1 ][line width=3]  (429,4287.43) -- (459,4287.43) -- (459,4317.18) -- (429,4317.18) -- cycle ;
\draw   (476,4227.93) -- (506,4227.93) -- (506,4257.68) -- (476,4257.68) -- cycle ;
\draw   (476,4257.68) -- (506,4257.68) -- (506,4287.43) -- (476,4287.43) -- cycle ;
\draw   (476,4227.93) -- (506,4227.93) -- (506,4257.68) -- (476,4257.68) -- cycle ;
\draw   (476,4317.18) -- (506,4317.18) -- (506,4346.93) -- (476,4346.93) -- cycle ;
\draw   (476,4346.93) -- (506,4346.93) -- (506,4376.68) -- (476,4376.68) -- cycle ;
\draw   (476,4346.93) -- (506,4346.93) -- (506,4376.68) -- (476,4376.68) -- cycle ;
\draw   (476,4406.43) -- (506,4406.43) -- (506,4436.18) -- (476,4436.18) -- cycle ;
\draw   (476,4436.18) -- (506,4436.18) -- (506,4465.93) -- (476,4465.93) -- cycle ;
\draw  [color={rgb, 255:red, 65; green, 117; blue, 5 }  ,draw opacity=1 ][line width=3]  (476,4465.93) -- (506,4465.93) -- (506,4495.68) -- (476,4495.68) -- cycle ;
\draw  [color={rgb, 255:red, 65; green, 117; blue, 5 }  ,draw opacity=1 ][line width=3]  (476,4376.68) -- (506,4376.68) -- (506,4406.43) -- (476,4406.43) -- cycle ;
\draw  [color={rgb, 255:red, 65; green, 117; blue, 5 }  ,draw opacity=1 ][line width=3]  (476,4287.43) -- (506,4287.43) -- (506,4317.18) -- (476,4317.18) -- cycle ;
\draw   (523,4227.93) -- (553,4227.93) -- (553,4257.68) -- (523,4257.68) -- cycle ;
\draw   (523,4257.68) -- (553,4257.68) -- (553,4287.43) -- (523,4287.43) -- cycle ;
\draw   (523,4227.93) -- (553,4227.93) -- (553,4257.68) -- (523,4257.68) -- cycle ;
\draw   (523,4317.18) -- (553,4317.18) -- (553,4346.93) -- (523,4346.93) -- cycle ;
\draw   (523,4346.93) -- (553,4346.93) -- (553,4376.68) -- (523,4376.68) -- cycle ;
\draw   (523,4346.93) -- (553,4346.93) -- (553,4376.68) -- (523,4376.68) -- cycle ;
\draw   (523,4406.43) -- (553,4406.43) -- (553,4436.18) -- (523,4436.18) -- cycle ;
\draw   (523,4436.18) -- (553,4436.18) -- (553,4465.93) -- (523,4465.93) -- cycle ;
\draw  [color={rgb, 255:red, 65; green, 117; blue, 5 }  ,draw opacity=1 ][line width=3]  (523,4198.18) -- (553,4198.18) -- (553,4227.93) -- (523,4227.93) -- cycle ;
\draw  [color={rgb, 255:red, 65; green, 117; blue, 5 }  ,draw opacity=1 ][line width=3]  (523,4465.93) -- (553,4465.93) -- (553,4495.68) -- (523,4495.68) -- cycle ;
\draw  [color={rgb, 255:red, 65; green, 117; blue, 5 }  ,draw opacity=1 ][line width=3]  (523,4376.68) -- (553,4376.68) -- (553,4406.43) -- (523,4406.43) -- cycle ;
\draw  [fill={rgb, 255:red, 155; green, 155; blue, 155 }  ,fill opacity=1 ] (429,4406.43) -- (459,4406.43) -- (459,4436.18) -- (429,4436.18) -- cycle ;
\draw  [fill={rgb, 255:red, 155; green, 155; blue, 155 }  ,fill opacity=1 ] (476,4227.93) -- (506,4227.93) -- (506,4257.68) -- (476,4257.68) -- cycle ;
\draw  [fill={rgb, 255:red, 155; green, 155; blue, 155 }  ,fill opacity=1 ] (523,4317.18) -- (553,4317.18) -- (553,4346.93) -- (523,4346.93) -- cycle ;
\draw  [color={rgb, 255:red, 65; green, 117; blue, 5 }  ,draw opacity=1 ][line width=3]  (429,4376.68) -- (459,4376.68) -- (459,4406.43) -- (429,4406.43) -- cycle ;
\draw  [color={rgb, 255:red, 65; green, 117; blue, 5 }  ,draw opacity=1 ][line width=3]  (523,4287.43) -- (553,4287.43) -- (553,4317.18) -- (523,4317.18) -- cycle ;
\draw  [color={rgb, 255:red, 65; green, 117; blue, 5 }  ,draw opacity=1 ][line width=3]  (476,4198.18) -- (506,4198.18) -- (506,4227.93) -- (476,4227.93) -- cycle ;

\draw (393,4467.43) node [anchor=north west][inner sep=0.75pt]   [align=left] {{\fontfamily{pcr}\selectfont 0}};
\draw (393,4439.43) node [anchor=north west][inner sep=0.75pt]   [align=left] {{\fontfamily{pcr}\selectfont 1}};
\draw (393,4409.43) node [anchor=north west][inner sep=0.75pt]  [color={rgb, 255:red, 0; green, 0; blue, 0 }  ,opacity=1 ] [align=left] {{\fontfamily{pcr}\selectfont 2}};
\draw (393,4378.43) node [anchor=north west][inner sep=0.75pt]   [align=left] {{\fontfamily{pcr}\selectfont 3}};
\draw (393,4347.43) node [anchor=north west][inner sep=0.75pt]   [align=left] {{\fontfamily{pcr}\selectfont 4}};
\draw (393,4320.43) node [anchor=north west][inner sep=0.75pt]   [align=left] {{\fontfamily{pcr}\selectfont 5}};
\draw (393,4291.43) node [anchor=north west][inner sep=0.75pt]   [align=left] {{\fontfamily{pcr}\selectfont 6}};
\draw (392,4263.43) node [anchor=north west][inner sep=0.75pt]   [align=left] {{\fontfamily{pcr}\selectfont 7}};
\draw (392,4233.43) node [anchor=north west][inner sep=0.75pt]   [align=left] {{\fontfamily{pcr}\selectfont 8}};
\draw (392,4202.43) node [anchor=north west][inner sep=0.75pt]   [align=left] {{\fontfamily{pcr}\selectfont 9}};
\draw (437,4469.43) node [anchor=north west][inner sep=0.75pt]   [align=left] {{\fontfamily{pcr}\selectfont 0}};
\draw (436,4379.43) node [anchor=north west][inner sep=0.75pt]   [align=left] {{\fontfamily{pcr}\selectfont 0}};
\draw (436,4291.43) node [anchor=north west][inner sep=0.75pt]   [align=left] {{\fontfamily{pcr}\selectfont 0}};
\draw (436,4201.43) node [anchor=north west][inner sep=0.75pt]   [align=left] {{\fontfamily{pcr}\selectfont 0}};
\draw (437,4439.43) node [anchor=north west][inner sep=0.75pt]   [align=left] {{\fontfamily{pcr}\selectfont 1}};
\draw (436,4349.43) node [anchor=north west][inner sep=0.75pt]   [align=left] {{\fontfamily{pcr}\selectfont 1}};
\draw (436,4261.43) node [anchor=north west][inner sep=0.75pt]   [align=left] {{\fontfamily{pcr}\selectfont 1}};
\draw (438,4410.43) node [anchor=north west][inner sep=0.75pt]   [align=left] {{\fontfamily{pcr}\selectfont 2}};
\draw (436,4321.43) node [anchor=north west][inner sep=0.75pt]   [align=left] {{\fontfamily{pcr}\selectfont 2}};
\draw (436,4232.43) node [anchor=north west][inner sep=0.75pt]   [align=left] {{\fontfamily{pcr}\selectfont 2}};
\draw (484,4469.43) node [anchor=north west][inner sep=0.75pt]   [align=left] {{\fontfamily{pcr}\selectfont 0}};
\draw (483,4379.43) node [anchor=north west][inner sep=0.75pt]   [align=left] {{\fontfamily{pcr}\selectfont 0}};
\draw (483,4291.43) node [anchor=north west][inner sep=0.75pt]   [align=left] {{\fontfamily{pcr}\selectfont 0}};
\draw (483,4201.43) node [anchor=north west][inner sep=0.75pt]   [align=left] {{\fontfamily{pcr}\selectfont 0}};
\draw (484,4439.43) node [anchor=north west][inner sep=0.75pt]   [align=left] {{\fontfamily{pcr}\selectfont 1}};
\draw (483,4349.43) node [anchor=north west][inner sep=0.75pt]   [align=left] {{\fontfamily{pcr}\selectfont 1}};
\draw (483,4261.43) node [anchor=north west][inner sep=0.75pt]   [align=left] {{\fontfamily{pcr}\selectfont 1}};
\draw (485,4410.43) node [anchor=north west][inner sep=0.75pt]   [align=left] {{\fontfamily{pcr}\selectfont 2}};
\draw (483,4321.43) node [anchor=north west][inner sep=0.75pt]   [align=left] {{\fontfamily{pcr}\selectfont 2}};
\draw (483,4232.43) node [anchor=north west][inner sep=0.75pt]   [align=left] {{\fontfamily{pcr}\selectfont 2}};
\draw (531,4469.43) node [anchor=north west][inner sep=0.75pt]   [align=left] {{\fontfamily{pcr}\selectfont 0}};
\draw (530,4379.43) node [anchor=north west][inner sep=0.75pt]   [align=left] {{\fontfamily{pcr}\selectfont 0}};
\draw (530,4291.43) node [anchor=north west][inner sep=0.75pt]   [align=left] {{\fontfamily{pcr}\selectfont 0}};
\draw (530,4201.43) node [anchor=north west][inner sep=0.75pt]   [align=left] {{\fontfamily{pcr}\selectfont 0}};
\draw (531,4439.43) node [anchor=north west][inner sep=0.75pt]   [align=left] {{\fontfamily{pcr}\selectfont 1}};
\draw (530,4349.43) node [anchor=north west][inner sep=0.75pt]   [align=left] {{\fontfamily{pcr}\selectfont 1}};
\draw (530,4261.43) node [anchor=north west][inner sep=0.75pt]   [align=left] {{\fontfamily{pcr}\selectfont 1}};
\draw (532,4410.43) node [anchor=north west][inner sep=0.75pt]   [align=left] {{\fontfamily{pcr}\selectfont 2}};
\draw (530,4321.43) node [anchor=north west][inner sep=0.75pt]   [align=left] {{\fontfamily{pcr}\selectfont 2}};
\draw (530,4232.43) node [anchor=north west][inner sep=0.75pt]   [align=left] {{\fontfamily{pcr}\selectfont 2}};

\end{tikzpicture}

%% file: figures/4lvl_2axis_no_info.tex
\tikzset{every picture/.style={line width=0.75pt}} 

\begin{tikzpicture}[x=0.75pt,y=0.75pt,yscale=-1,xscale=1]

\draw   (376.38,2593.59) -- (376.32,2623.59) -- (197.75,2623.25) -- (197.81,2593.25) -- cycle ;
\draw  [color={rgb, 255:red, 65; green, 117; blue, 5 }  ,draw opacity=1 ][line width=2.25]  (376.43,2563.59) -- (376.38,2593.59) -- (197.94,2593.25) -- (198,2563.25) -- cycle ;
\draw   (376.45,2623.59) -- (376.39,2653.59) -- (197.83,2653.25) -- (197.88,2623.25) -- cycle ;
\draw  [color={rgb, 255:red, 65; green, 117; blue, 5 }  ,draw opacity=1 ][line width=2.25]  (376.44,2653.59) -- (376.38,2683.59) -- (197.77,2683.25) -- (197.83,2653.25) -- cycle ;
\draw    (302,2578) -- (354.42,2578.24) ;
\draw [shift={(356.42,2578.25)}, rotate = 180.26] [color={rgb, 255:red, 0; green, 0; blue, 0 }  ][line width=0.75]    (10.93,-3.29) .. controls (6.95,-1.4) and (3.31,-0.3) .. (0,0) .. controls (3.31,0.3) and (6.95,1.4) .. (10.93,3.29)   ;
\draw    (300,2668) -- (352.42,2668.24) ;
\draw [shift={(354.42,2668.25)}, rotate = 180.26] [color={rgb, 255:red, 0; green, 0; blue, 0 }  ][line width=0.75]    (10.93,-3.29) .. controls (6.95,-1.4) and (3.31,-0.3) .. (0,0) .. controls (3.31,0.3) and (6.95,1.4) .. (10.93,3.29)   ;
\draw    (265.42,2667.25) -- (217.42,2667.25) ;
\draw [shift={(215.42,2667.25)}, rotate = 360] [color={rgb, 255:red, 0; green, 0; blue, 0 }  ][line width=0.75]    (10.93,-3.29) .. controls (6.95,-1.4) and (3.31,-0.3) .. (0,0) .. controls (3.31,0.3) and (6.95,1.4) .. (10.93,3.29)   ;
\draw    (263.42,2577.25) -- (215.42,2577.25) ;
\draw [shift={(213.42,2577.25)}, rotate = 360] [color={rgb, 255:red, 0; green, 0; blue, 0 }  ][line width=0.75]    (10.93,-3.29) .. controls (6.95,-1.4) and (3.31,-0.3) .. (0,0) .. controls (3.31,0.3) and (6.95,1.4) .. (10.93,3.29)   ;
\draw    (303,2610) -- (355.42,2610.24) ;
\draw [shift={(357.42,2610.25)}, rotate = 180.26] [color={rgb, 255:red, 0; green, 0; blue, 0 }  ][line width=0.75]    (10.93,-3.29) .. controls (6.95,-1.4) and (3.31,-0.3) .. (0,0) .. controls (3.31,0.3) and (6.95,1.4) .. (10.93,3.29)   ;
\draw    (264.42,2609.25) -- (216.42,2609.25) ;
\draw [shift={(214.42,2609.25)}, rotate = 360] [color={rgb, 255:red, 0; green, 0; blue, 0 }  ][line width=0.75]    (10.93,-3.29) .. controls (6.95,-1.4) and (3.31,-0.3) .. (0,0) .. controls (3.31,0.3) and (6.95,1.4) .. (10.93,3.29)   ;
\draw    (304,2638) -- (356.42,2638.24) ;
\draw [shift={(358.42,2638.25)}, rotate = 180.26] [color={rgb, 255:red, 0; green, 0; blue, 0 }  ][line width=0.75]    (10.93,-3.29) .. controls (6.95,-1.4) and (3.31,-0.3) .. (0,0) .. controls (3.31,0.3) and (6.95,1.4) .. (10.93,3.29)   ;
\draw    (265.42,2637.25) -- (217.42,2637.25) ;
\draw [shift={(215.42,2637.25)}, rotate = 360] [color={rgb, 255:red, 0; green, 0; blue, 0 }  ][line width=0.75]    (10.93,-3.29) .. controls (6.95,-1.4) and (3.31,-0.3) .. (0,0) .. controls (3.31,0.3) and (6.95,1.4) .. (10.93,3.29)   ;

\draw (178,2656.14) node [anchor=north west][inner sep=0.75pt]   [align=left] {{\fontfamily{pcr}\selectfont 0}};
\draw (179,2626.14) node [anchor=north west][inner sep=0.75pt]   [align=left] {{\fontfamily{pcr}\selectfont 1}};
\draw (180,2596.14) node [anchor=north west][inner sep=0.75pt]   [align=left] {{\fontfamily{pcr}\selectfont 2}};
\draw (180,2567.14) node [anchor=north west][inner sep=0.75pt]   [align=left] {{\fontfamily{pcr}\selectfont 3}};
\draw (207,2684.14) node [anchor=north west][inner sep=0.75pt]   [align=left] {{\fontfamily{pcr}\selectfont 0}};
\draw (239,2684.14) node [anchor=north west][inner sep=0.75pt]   [align=left] {{\fontfamily{pcr}\selectfont 1}};
\draw (267,2684.14) node [anchor=north west][inner sep=0.75pt]   [align=left] {{\fontfamily{pcr}\selectfont 2}};
\draw (295,2684.14) node [anchor=north west][inner sep=0.75pt]   [align=left] {{\fontfamily{pcr}\selectfont 3}};
\draw (324,2684.14) node [anchor=north west][inner sep=0.75pt]   [align=left] {{\fontfamily{pcr}\selectfont 4}};
\draw (353,2684.14) node [anchor=north west][inner sep=0.75pt]   [align=left] {{\fontfamily{pcr}\selectfont 5}};
\draw (275,2656.14) node [anchor=north west][inner sep=0.75pt]   [align=left] {{\fontfamily{pcr}\selectfont \textbf{?}}};
\draw (274,2566.14) node [anchor=north west][inner sep=0.75pt]   [align=left] {{\fontfamily{pcr}\selectfont \textbf{?}}};
\draw (275,2598.14) node [anchor=north west][inner sep=0.75pt]   [align=left] {{\fontfamily{pcr}\selectfont \textbf{?}}};
\draw (276,2626.14) node [anchor=north west][inner sep=0.75pt]   [align=left] {{\fontfamily{pcr}\selectfont \textbf{?}}};

\end{tikzpicture}

%% file: figures/10lvl_3multiples.tex
\tikzset{every picture/.style={line width=0.75pt}} 

\begin{tikzpicture}[x=0.75pt,y=0.75pt,yscale=-1,xscale=1]

\draw   (25,3133) -- (55,3133) -- (55,3162.75) -- (25,3162.75) -- cycle ;
\draw   (25,3162.75) -- (55,3162.75) -- (55,3192.5) -- (25,3192.5) -- cycle ;
\draw   (25,3133) -- (55,3133) -- (55,3162.75) -- (25,3162.75) -- cycle ;
\draw   (25,3222.25) -- (55,3222.25) -- (55,3252) -- (25,3252) -- cycle ;
\draw   (25,3252) -- (55,3252) -- (55,3281.75) -- (25,3281.75) -- cycle ;
\draw   (25,3252) -- (55,3252) -- (55,3281.75) -- (25,3281.75) -- cycle ;
\draw   (25,3311.5) -- (55,3311.5) -- (55,3341.25) -- (25,3341.25) -- cycle ;
\draw   (25,3341.25) -- (55,3341.25) -- (55,3371) -- (25,3371) -- cycle ;
\draw  [color={rgb, 255:red, 65; green, 117; blue, 5 }  ,draw opacity=1 ][line width=2.25]  (25,3103.25) -- (55,3103.25) -- (55,3133) -- (25,3133) -- cycle ;
\draw  [color={rgb, 255:red, 65; green, 117; blue, 5 }  ,draw opacity=1 ][line width=2.25]  (25,3371) -- (55,3371) -- (55,3400.75) -- (25,3400.75) -- cycle ;
\draw  [color={rgb, 255:red, 65; green, 117; blue, 5 }  ,draw opacity=1 ][line width=2.25]  (25,3281.75) -- (55,3281.75) -- (55,3311.5) -- (25,3311.5) -- cycle ;
\draw  [color={rgb, 255:red, 65; green, 117; blue, 5 }  ,draw opacity=1 ][line width=2.25]  (25,3192.5) -- (55,3192.5) -- (55,3222.25) -- (25,3222.25) -- cycle ;

\draw (8,3372.5) node [anchor=north west][inner sep=0.75pt]   [align=left] {{\fontfamily{pcr}\selectfont 0}};
\draw (9,3344.5) node [anchor=north west][inner sep=0.75pt]   [align=left] {{\fontfamily{pcr}\selectfont 1}};
\draw (8,3314.5) node [anchor=north west][inner sep=0.75pt]   [align=left] {{\fontfamily{pcr}\selectfont 2}};
\draw (8,3283.5) node [anchor=north west][inner sep=0.75pt]   [align=left] {{\fontfamily{pcr}\selectfont 3}};
\draw (8,3252.5) node [anchor=north west][inner sep=0.75pt]   [align=left] {{\fontfamily{pcr}\selectfont 4}};
\draw (8,3225.5) node [anchor=north west][inner sep=0.75pt]   [align=left] {{\fontfamily{pcr}\selectfont 5}};
\draw (8,3196.5) node [anchor=north west][inner sep=0.75pt]   [align=left] {{\fontfamily{pcr}\selectfont 6}};
\draw (7,3168.5) node [anchor=north west][inner sep=0.75pt]   [align=left] {{\fontfamily{pcr}\selectfont 7}};
\draw (7,3138.5) node [anchor=north west][inner sep=0.75pt]   [align=left] {{\fontfamily{pcr}\selectfont 8}};
\draw (7,3107.5) node [anchor=north west][inner sep=0.75pt]   [align=left] {{\fontfamily{pcr}\selectfont 9}};
\draw (25,3287.5) node [anchor=north west][inner sep=0.75pt]  [font=\footnotesize] [align=left] {$\displaystyle \ket{1_{L}}$};
\draw (26,3108.5) node [anchor=north west][inner sep=0.75pt]  [font=\footnotesize] [align=left] {$\displaystyle \ket{1_{L}}$};
\draw (26,3198.5) node [anchor=north west][inner sep=0.75pt]  [font=\footnotesize] [align=left] {$\displaystyle \ket{0_{L}}$};
\draw (25,3376.5) node [anchor=north west][inner sep=0.75pt]  [font=\footnotesize] [align=left] {$\displaystyle \ket{0_{L}}$};

\end{tikzpicture}

%% file: figures/error_transitions_diagram.tex
\tikzset{every picture/.style={line width=0.75pt}} 

\begin{tikzpicture}[x=0.75pt,y=0.75pt,yscale=-1,xscale=1]

\draw   (261.33,5812.17) .. controls (261.33,5803.42) and (268.42,5796.33) .. (277.17,5796.33) .. controls (285.91,5796.33) and (293,5803.42) .. (293,5812.17) .. controls (293,5820.91) and (285.91,5828) .. (277.17,5828) .. controls (268.42,5828) and (261.33,5820.91) .. (261.33,5812.17) -- cycle ;
\draw   (348.33,5813.17) .. controls (348.33,5804.42) and (355.42,5797.33) .. (364.17,5797.33) .. controls (372.91,5797.33) and (380,5804.42) .. (380,5813.17) .. controls (380,5821.91) and (372.91,5829) .. (364.17,5829) .. controls (355.42,5829) and (348.33,5821.91) .. (348.33,5813.17) -- cycle ;
\draw   (263.33,5741.17) .. controls (263.33,5732.42) and (270.42,5725.33) .. (279.17,5725.33) .. controls (287.91,5725.33) and (295,5732.42) .. (295,5741.17) .. controls (295,5749.91) and (287.91,5757) .. (279.17,5757) .. controls (270.42,5757) and (263.33,5749.91) .. (263.33,5741.17) -- cycle ;
\draw   (351.33,5742.17) .. controls (351.33,5733.42) and (358.42,5726.33) .. (367.17,5726.33) .. controls (375.91,5726.33) and (383,5733.42) .. (383,5742.17) .. controls (383,5750.91) and (375.91,5758) .. (367.17,5758) .. controls (358.42,5758) and (351.33,5750.91) .. (351.33,5742.17) -- cycle ;
\draw [color={rgb, 255:red, 224; green, 16; blue, 218 }  ,draw opacity=1 ]   (277.17,5796.33) -- (279.07,5759) ;
\draw [shift={(279.17,5757)}, rotate = 92.91] [color={rgb, 255:red, 224; green, 16; blue, 218 }  ,draw opacity=1 ][line width=0.75]    (10.93,-3.29) .. controls (6.95,-1.4) and (3.31,-0.3) .. (0,0) .. controls (3.31,0.3) and (6.95,1.4) .. (10.93,3.29)   ;
\draw [color={rgb, 255:red, 56; green, 195; blue, 226 }  ,draw opacity=1 ]   (293,5812.17) -- (346.33,5813.13) ;
\draw [shift={(348.33,5813.17)}, rotate = 181.04] [color={rgb, 255:red, 56; green, 195; blue, 226 }  ,draw opacity=1 ][line width=0.75]    (10.93,-3.29) .. controls (6.95,-1.4) and (3.31,-0.3) .. (0,0) .. controls (3.31,0.3) and (6.95,1.4) .. (10.93,3.29)   ;
\draw [color={rgb, 255:red, 56; green, 195; blue, 226 }  ,draw opacity=1 ]   (295,5741.17) -- (349.33,5742.13) ;
\draw [shift={(351.33,5742.17)}, rotate = 181.02] [color={rgb, 255:red, 56; green, 195; blue, 226 }  ,draw opacity=1 ][line width=0.75]    (10.93,-3.29) .. controls (6.95,-1.4) and (3.31,-0.3) .. (0,0) .. controls (3.31,0.3) and (6.95,1.4) .. (10.93,3.29)   ;
\draw [color={rgb, 255:red, 56; green, 195; blue, 226 }  ,draw opacity=1 ]   (367.17,5726.33) .. controls (347.2,5715.11) and (326.42,5705.2) .. (280.56,5724.73) ;
\draw [shift={(279.17,5725.33)}, rotate = 336.53] [color={rgb, 255:red, 56; green, 195; blue, 226 }  ,draw opacity=1 ][line width=0.75]    (10.93,-3.29) .. controls (6.95,-1.4) and (3.31,-0.3) .. (0,0) .. controls (3.31,0.3) and (6.95,1.4) .. (10.93,3.29)   ;
\draw [color={rgb, 255:red, 56; green, 195; blue, 226 }  ,draw opacity=1 ]   (364.17,5829) .. controls (332.64,5845.66) and (307.04,5845.03) .. (278.89,5829) ;
\draw [shift={(277.17,5828)}, rotate = 30.52] [color={rgb, 255:red, 56; green, 195; blue, 226 }  ,draw opacity=1 ][line width=0.75]    (10.93,-3.29) .. controls (6.95,-1.4) and (3.31,-0.3) .. (0,0) .. controls (3.31,0.3) and (6.95,1.4) .. (10.93,3.29)   ;
\draw [color={rgb, 255:red, 224; green, 16; blue, 218 }  ,draw opacity=1 ]   (263.33,5741.17) .. controls (253.36,5766.1) and (253.94,5793.97) .. (260.59,5810.42) ;
\draw [shift={(261.33,5812.17)}, rotate = 245.6] [color={rgb, 255:red, 224; green, 16; blue, 218 }  ,draw opacity=1 ][line width=0.75]    (10.93,-3.29) .. controls (6.95,-1.4) and (3.31,-0.3) .. (0,0) .. controls (3.31,0.3) and (6.95,1.4) .. (10.93,3.29)   ;

\draw (279.17,5757) node [anchor=north west][inner sep=0.75pt]  [font=\Large,color={rgb, 255:red, 224; green, 16; blue, 218 }  ,opacity=1 ]  {$x$};
\draw (221,5783) node [anchor=north west][inner sep=0.75pt]  [font=\Large,color={rgb, 255:red, 224; green, 16; blue, 218 }  ,opacity=1 ]  {$x^{-1}$};
\draw (324,5731) node [anchor=north west][inner sep=0.75pt]  [font=\Large,color={rgb, 255:red, 56; green, 195; blue, 226 }  ,opacity=1 ]  {$z$};
\draw (284,5687) node [anchor=north west][inner sep=0.75pt]  [font=\Large,color={rgb, 255:red, 56; green, 195; blue, 226 }  ,opacity=1 ]  {$z^{-1}$};
\draw (276,5834.98) node [anchor=north west][inner sep=0.75pt]  [font=\Large,color={rgb, 255:red, 56; green, 195; blue, 226 }  ,opacity=1 ]  {$z^{-1}$};
\draw (326,5782) node [anchor=north west][inner sep=0.75pt]  [font=\Large,color={rgb, 255:red, 56; green, 195; blue, 226 }  ,opacity=1 ]  {$z$};
\draw (263,5801.33) node [anchor=north west][inner sep=0.75pt]  [font=\scriptsize] [align=left] {$\displaystyle \ket{0_{L}}$};
\draw (349,5803.33) node [anchor=north west][inner sep=0.75pt]  [font=\scriptsize] [align=left] {$\displaystyle \ket{0_{L}}$};
\draw (264,5730.33) node [anchor=north west][inner sep=0.75pt]  [font=\scriptsize] [align=left] {$\displaystyle \ket{1_{L}}$};
\draw (350,5731.33) node [anchor=north west][inner sep=0.75pt]  [font=\scriptsize] [align=left] {$\displaystyle -\ket{1_{L}}$};

\end{tikzpicture}

%% file: figures/10x13lvl_full_grid.tex
\resizebox{0.5\textwidth}{!}{%

\tikzset{every picture/.style={line width=0.75pt}} 

\begin{tikzpicture}[x=0.75pt,y=0.75pt,yscale=-1,xscale=1]

\draw   (138,3132) -- (168,3132) -- (168,3161.75) -- (138,3161.75) -- cycle ;
\draw   (138,3161.75) -- (168,3161.75) -- (168,3191.5) -- (138,3191.5) -- cycle ;
\draw   (138,3132) -- (168,3132) -- (168,3161.75) -- (138,3161.75) -- cycle ;
\draw   (138,3221.25) -- (168,3221.25) -- (168,3251) -- (138,3251) -- cycle ;
\draw   (138,3251) -- (168,3251) -- (168,3280.75) -- (138,3280.75) -- cycle ;
\draw   (138,3251) -- (168,3251) -- (168,3280.75) -- (138,3280.75) -- cycle ;
\draw   (138,3310.5) -- (168,3310.5) -- (168,3340.25) -- (138,3340.25) -- cycle ;
\draw   (138,3340.25) -- (168,3340.25) -- (168,3370) -- (138,3370) -- cycle ;
\draw   (168,3132) -- (198,3132) -- (198,3161.75) -- (168,3161.75) -- cycle ;
\draw   (168,3161.75) -- (198,3161.75) -- (198,3191.5) -- (168,3191.5) -- cycle ;
\draw   (168,3132) -- (198,3132) -- (198,3161.75) -- (168,3161.75) -- cycle ;
\draw   (168,3221.25) -- (198,3221.25) -- (198,3251) -- (168,3251) -- cycle ;
\draw   (168,3251) -- (198,3251) -- (198,3280.75) -- (168,3280.75) -- cycle ;
\draw   (168,3251) -- (198,3251) -- (198,3280.75) -- (168,3280.75) -- cycle ;
\draw   (168,3310.5) -- (198,3310.5) -- (198,3340.25) -- (168,3340.25) -- cycle ;
\draw   (168,3340.25) -- (198,3340.25) -- (198,3370) -- (168,3370) -- cycle ;
\draw  [color={rgb, 255:red, 0; green, 0; blue, 0 }  ,draw opacity=1 ][line width=0.75]  (168,3102.25) -- (198,3102.25) -- (198,3132) -- (168,3132) -- cycle ;
\draw  [color={rgb, 255:red, 0; green, 0; blue, 0 }  ,draw opacity=1 ][line width=0.75]  (168,3370) -- (198,3370) -- (198,3399.75) -- (168,3399.75) -- cycle ;
\draw  [color={rgb, 255:red, 0; green, 0; blue, 0 }  ,draw opacity=1 ][line width=0.75]  (168,3280.75) -- (198,3280.75) -- (198,3310.5) -- (168,3310.5) -- cycle ;
\draw  [color={rgb, 255:red, 0; green, 0; blue, 0 }  ,draw opacity=1 ][line width=0.75]  (168,3191.5) -- (198,3191.5) -- (198,3221.25) -- (168,3221.25) -- cycle ;
\draw   (198,3132) -- (228,3132) -- (228,3161.75) -- (198,3161.75) -- cycle ;
\draw   (198,3161.75) -- (228,3161.75) -- (228,3191.5) -- (198,3191.5) -- cycle ;
\draw   (198,3132) -- (228,3132) -- (228,3161.75) -- (198,3161.75) -- cycle ;
\draw   (198,3221.25) -- (228,3221.25) -- (228,3251) -- (198,3251) -- cycle ;
\draw   (198,3251) -- (228,3251) -- (228,3280.75) -- (198,3280.75) -- cycle ;
\draw   (198,3251) -- (228,3251) -- (228,3280.75) -- (198,3280.75) -- cycle ;
\draw   (198,3310.5) -- (228,3310.5) -- (228,3340.25) -- (198,3340.25) -- cycle ;
\draw   (198,3340.25) -- (228,3340.25) -- (228,3370) -- (198,3370) -- cycle ;
\draw  [color={rgb, 255:red, 0; green, 0; blue, 0 }  ,draw opacity=1 ][line width=0.75]  (198,3102.25) -- (228,3102.25) -- (228,3132) -- (198,3132) -- cycle ;
\draw  [color={rgb, 255:red, 0; green, 0; blue, 0 }  ,draw opacity=1 ][line width=0.75]  (198,3370) -- (228,3370) -- (228,3399.75) -- (198,3399.75) -- cycle ;
\draw  [color={rgb, 255:red, 0; green, 0; blue, 0 }  ,draw opacity=1 ][line width=0.75]  (198,3280.75) -- (228,3280.75) -- (228,3310.5) -- (198,3310.5) -- cycle ;
\draw  [color={rgb, 255:red, 0; green, 0; blue, 0 }  ,draw opacity=1 ][line width=0.75]  (198,3191.5) -- (228,3191.5) -- (228,3221.25) -- (198,3221.25) -- cycle ;
\draw   (228,3132) -- (258,3132) -- (258,3161.75) -- (228,3161.75) -- cycle ;
\draw   (228,3161.75) -- (258,3161.75) -- (258,3191.5) -- (228,3191.5) -- cycle ;
\draw   (228,3132) -- (258,3132) -- (258,3161.75) -- (228,3161.75) -- cycle ;
\draw   (228,3221.25) -- (258,3221.25) -- (258,3251) -- (228,3251) -- cycle ;
\draw   (228,3251) -- (258,3251) -- (258,3280.75) -- (228,3280.75) -- cycle ;
\draw   (228,3251) -- (258,3251) -- (258,3280.75) -- (228,3280.75) -- cycle ;
\draw   (228,3310.5) -- (258,3310.5) -- (258,3340.25) -- (228,3340.25) -- cycle ;
\draw   (228,3340.25) -- (258,3340.25) -- (258,3370) -- (228,3370) -- cycle ;
\draw  [color={rgb, 255:red, 0; green, 0; blue, 0 }  ,draw opacity=1 ][line width=0.75]  (228,3102.25) -- (258,3102.25) -- (258,3132) -- (228,3132) -- cycle ;
\draw  [color={rgb, 255:red, 0; green, 0; blue, 0 }  ,draw opacity=1 ][line width=0.75]  (228,3370) -- (258,3370) -- (258,3399.75) -- (228,3399.75) -- cycle ;
\draw  [color={rgb, 255:red, 0; green, 0; blue, 0 }  ,draw opacity=1 ][line width=0.75]  (228,3280.75) -- (258,3280.75) -- (258,3310.5) -- (228,3310.5) -- cycle ;
\draw  [color={rgb, 255:red, 0; green, 0; blue, 0 }  ,draw opacity=1 ][line width=0.75]  (228,3191.5) -- (258,3191.5) -- (258,3221.25) -- (228,3221.25) -- cycle ;
\draw   (258,3132) -- (288,3132) -- (288,3161.75) -- (258,3161.75) -- cycle ;
\draw   (258,3161.75) -- (288,3161.75) -- (288,3191.5) -- (258,3191.5) -- cycle ;
\draw   (258,3132) -- (288,3132) -- (288,3161.75) -- (258,3161.75) -- cycle ;
\draw   (258,3221.25) -- (288,3221.25) -- (288,3251) -- (258,3251) -- cycle ;
\draw   (258,3251) -- (288,3251) -- (288,3280.75) -- (258,3280.75) -- cycle ;
\draw   (258,3251) -- (288,3251) -- (288,3280.75) -- (258,3280.75) -- cycle ;
\draw   (258,3310.5) -- (288,3310.5) -- (288,3340.25) -- (258,3340.25) -- cycle ;
\draw   (258,3340.25) -- (288,3340.25) -- (288,3370) -- (258,3370) -- cycle ;
\draw   (288,3132) -- (318,3132) -- (318,3161.75) -- (288,3161.75) -- cycle ;
\draw   (288,3161.75) -- (318,3161.75) -- (318,3191.5) -- (288,3191.5) -- cycle ;
\draw   (288,3132) -- (318,3132) -- (318,3161.75) -- (288,3161.75) -- cycle ;
\draw   (288,3221.25) -- (318,3221.25) -- (318,3251) -- (288,3251) -- cycle ;
\draw   (288,3251) -- (318,3251) -- (318,3280.75) -- (288,3280.75) -- cycle ;
\draw   (288,3251) -- (318,3251) -- (318,3280.75) -- (288,3280.75) -- cycle ;
\draw   (288,3310.5) -- (318,3310.5) -- (318,3340.25) -- (288,3340.25) -- cycle ;
\draw   (288,3340.25) -- (318,3340.25) -- (318,3370) -- (288,3370) -- cycle ;
\draw  [color={rgb, 255:red, 0; green, 0; blue, 0 }  ,draw opacity=1 ][line width=0.75]  (288,3102.25) -- (318,3102.25) -- (318,3132) -- (288,3132) -- cycle ;
\draw  [color={rgb, 255:red, 0; green, 0; blue, 0 }  ,draw opacity=1 ][line width=0.75]  (288,3370) -- (318,3370) -- (318,3399.75) -- (288,3399.75) -- cycle ;
\draw  [color={rgb, 255:red, 0; green, 0; blue, 0 }  ,draw opacity=1 ][line width=0.75]  (288,3280.75) -- (318,3280.75) -- (318,3310.5) -- (288,3310.5) -- cycle ;
\draw  [color={rgb, 255:red, 0; green, 0; blue, 0 }  ,draw opacity=1 ][line width=0.75]  (288,3191.5) -- (318,3191.5) -- (318,3221.25) -- (288,3221.25) -- cycle ;
\draw   (318,3132) -- (348,3132) -- (348,3161.75) -- (318,3161.75) -- cycle ;
\draw   (318,3161.75) -- (348,3161.75) -- (348,3191.5) -- (318,3191.5) -- cycle ;
\draw   (318,3132) -- (348,3132) -- (348,3161.75) -- (318,3161.75) -- cycle ;
\draw   (318,3221.25) -- (348,3221.25) -- (348,3251) -- (318,3251) -- cycle ;
\draw   (318,3251) -- (348,3251) -- (348,3280.75) -- (318,3280.75) -- cycle ;
\draw   (318,3251) -- (348,3251) -- (348,3280.75) -- (318,3280.75) -- cycle ;
\draw   (318,3310.5) -- (348,3310.5) -- (348,3340.25) -- (318,3340.25) -- cycle ;
\draw   (318,3340.25) -- (348,3340.25) -- (348,3370) -- (318,3370) -- cycle ;
\draw  [color={rgb, 255:red, 0; green, 0; blue, 0 }  ,draw opacity=1 ][line width=0.75]  (318,3102.25) -- (348,3102.25) -- (348,3132) -- (318,3132) -- cycle ;
\draw  [color={rgb, 255:red, 0; green, 0; blue, 0 }  ,draw opacity=1 ][line width=0.75]  (318,3370) -- (348,3370) -- (348,3399.75) -- (318,3399.75) -- cycle ;
\draw  [color={rgb, 255:red, 0; green, 0; blue, 0 }  ,draw opacity=1 ][line width=0.75]  (318,3280.75) -- (348,3280.75) -- (348,3310.5) -- (318,3310.5) -- cycle ;
\draw  [color={rgb, 255:red, 0; green, 0; blue, 0 }  ,draw opacity=1 ][line width=0.75]  (318,3191.5) -- (348,3191.5) -- (348,3221.25) -- (318,3221.25) -- cycle ;
\draw   (348,3132) -- (378,3132) -- (378,3161.75) -- (348,3161.75) -- cycle ;
\draw   (348,3161.75) -- (378,3161.75) -- (378,3191.5) -- (348,3191.5) -- cycle ;
\draw   (348,3132) -- (378,3132) -- (378,3161.75) -- (348,3161.75) -- cycle ;
\draw   (348,3221.25) -- (378,3221.25) -- (378,3251) -- (348,3251) -- cycle ;
\draw   (348,3251) -- (378,3251) -- (378,3280.75) -- (348,3280.75) -- cycle ;
\draw   (348,3251) -- (378,3251) -- (378,3280.75) -- (348,3280.75) -- cycle ;
\draw   (348,3310.5) -- (378,3310.5) -- (378,3340.25) -- (348,3340.25) -- cycle ;
\draw   (348,3340.25) -- (378,3340.25) -- (378,3370) -- (348,3370) -- cycle ;
\draw  [color={rgb, 255:red, 0; green, 0; blue, 0 }  ,draw opacity=1 ][line width=0.75]  (348,3102.25) -- (378,3102.25) -- (378,3132) -- (348,3132) -- cycle ;
\draw  [color={rgb, 255:red, 0; green, 0; blue, 0 }  ,draw opacity=1 ][line width=0.75]  (348,3370) -- (378,3370) -- (378,3399.75) -- (348,3399.75) -- cycle ;
\draw  [color={rgb, 255:red, 0; green, 0; blue, 0 }  ,draw opacity=1 ][line width=0.75]  (348,3280.75) -- (378,3280.75) -- (378,3310.5) -- (348,3310.5) -- cycle ;
\draw  [color={rgb, 255:red, 0; green, 0; blue, 0 }  ,draw opacity=1 ][line width=0.75]  (348,3191.5) -- (378,3191.5) -- (378,3221.25) -- (348,3221.25) -- cycle ;
\draw   (378,3132) -- (408,3132) -- (408,3161.75) -- (378,3161.75) -- cycle ;
\draw   (378,3161.75) -- (408,3161.75) -- (408,3191.5) -- (378,3191.5) -- cycle ;
\draw   (378,3132) -- (408,3132) -- (408,3161.75) -- (378,3161.75) -- cycle ;
\draw   (378,3221.25) -- (408,3221.25) -- (408,3251) -- (378,3251) -- cycle ;
\draw   (378,3251) -- (408,3251) -- (408,3280.75) -- (378,3280.75) -- cycle ;
\draw   (378,3251) -- (408,3251) -- (408,3280.75) -- (378,3280.75) -- cycle ;
\draw   (378,3310.5) -- (408,3310.5) -- (408,3340.25) -- (378,3340.25) -- cycle ;
\draw   (378,3340.25) -- (408,3340.25) -- (408,3370) -- (378,3370) -- cycle ;
\draw   (408,3132) -- (438,3132) -- (438,3161.75) -- (408,3161.75) -- cycle ;
\draw   (408,3161.75) -- (438,3161.75) -- (438,3191.5) -- (408,3191.5) -- cycle ;
\draw   (408,3132) -- (438,3132) -- (438,3161.75) -- (408,3161.75) -- cycle ;
\draw   (408,3221.25) -- (438,3221.25) -- (438,3251) -- (408,3251) -- cycle ;
\draw   (408,3251) -- (438,3251) -- (438,3280.75) -- (408,3280.75) -- cycle ;
\draw   (408,3251) -- (438,3251) -- (438,3280.75) -- (408,3280.75) -- cycle ;
\draw   (408,3310.5) -- (438,3310.5) -- (438,3340.25) -- (408,3340.25) -- cycle ;
\draw   (408,3340.25) -- (438,3340.25) -- (438,3370) -- (408,3370) -- cycle ;
\draw  [color={rgb, 255:red, 0; green, 0; blue, 0 }  ,draw opacity=1 ][line width=0.75]  (408,3102.25) -- (438,3102.25) -- (438,3132) -- (408,3132) -- cycle ;
\draw  [color={rgb, 255:red, 0; green, 0; blue, 0 }  ,draw opacity=1 ][line width=0.75]  (408,3370) -- (438,3370) -- (438,3399.75) -- (408,3399.75) -- cycle ;
\draw  [color={rgb, 255:red, 0; green, 0; blue, 0 }  ,draw opacity=1 ][line width=0.75]  (408,3280.75) -- (438,3280.75) -- (438,3310.5) -- (408,3310.5) -- cycle ;
\draw  [color={rgb, 255:red, 0; green, 0; blue, 0 }  ,draw opacity=1 ][line width=0.75]  (408,3191.5) -- (438,3191.5) -- (438,3221.25) -- (408,3221.25) -- cycle ;
\draw   (438,3132) -- (468,3132) -- (468,3161.75) -- (438,3161.75) -- cycle ;
\draw   (438,3161.75) -- (468,3161.75) -- (468,3191.5) -- (438,3191.5) -- cycle ;
\draw   (438,3132) -- (468,3132) -- (468,3161.75) -- (438,3161.75) -- cycle ;
\draw   (438,3221.25) -- (468,3221.25) -- (468,3251) -- (438,3251) -- cycle ;
\draw   (438,3251) -- (468,3251) -- (468,3280.75) -- (438,3280.75) -- cycle ;
\draw   (438,3251) -- (468,3251) -- (468,3280.75) -- (438,3280.75) -- cycle ;
\draw   (438,3310.5) -- (468,3310.5) -- (468,3340.25) -- (438,3340.25) -- cycle ;
\draw   (438,3340.25) -- (468,3340.25) -- (468,3370) -- (438,3370) -- cycle ;
\draw  [color={rgb, 255:red, 0; green, 0; blue, 0 }  ,draw opacity=1 ][line width=0.75]  (438,3102.25) -- (468,3102.25) -- (468,3132) -- (438,3132) -- cycle ;
\draw  [color={rgb, 255:red, 0; green, 0; blue, 0 }  ,draw opacity=1 ][line width=0.75]  (438,3370) -- (468,3370) -- (468,3399.75) -- (438,3399.75) -- cycle ;
\draw  [color={rgb, 255:red, 0; green, 0; blue, 0 }  ,draw opacity=1 ][line width=0.75]  (438,3280.75) -- (468,3280.75) -- (468,3310.5) -- (438,3310.5) -- cycle ;
\draw  [color={rgb, 255:red, 0; green, 0; blue, 0 }  ,draw opacity=1 ][line width=0.75]  (438,3191.5) -- (468,3191.5) -- (468,3221.25) -- (438,3221.25) -- cycle ;
\draw   (468,3132) -- (498,3132) -- (498,3161.75) -- (468,3161.75) -- cycle ;
\draw   (468,3161.75) -- (498,3161.75) -- (498,3191.5) -- (468,3191.5) -- cycle ;
\draw   (468,3132) -- (498,3132) -- (498,3161.75) -- (468,3161.75) -- cycle ;
\draw   (468,3221.25) -- (498,3221.25) -- (498,3251) -- (468,3251) -- cycle ;
\draw   (468,3251) -- (498,3251) -- (498,3280.75) -- (468,3280.75) -- cycle ;
\draw   (468,3251) -- (498,3251) -- (498,3280.75) -- (468,3280.75) -- cycle ;
\draw   (468,3310.5) -- (498,3310.5) -- (498,3340.25) -- (468,3340.25) -- cycle ;
\draw   (468,3340.25) -- (498,3340.25) -- (498,3370) -- (468,3370) -- cycle ;
\draw  [color={rgb, 255:red, 0; green, 0; blue, 0 }  ,draw opacity=1 ][line width=0.75]  (468,3102.25) -- (498,3102.25) -- (498,3132) -- (468,3132) -- cycle ;
\draw  [color={rgb, 255:red, 0; green, 0; blue, 0 }  ,draw opacity=1 ][line width=0.75]  (468,3370) -- (498,3370) -- (498,3399.75) -- (468,3399.75) -- cycle ;
\draw  [color={rgb, 255:red, 0; green, 0; blue, 0 }  ,draw opacity=1 ][line width=0.75]  (468,3280.75) -- (498,3280.75) -- (498,3310.5) -- (468,3310.5) -- cycle ;
\draw  [color={rgb, 255:red, 0; green, 0; blue, 0 }  ,draw opacity=1 ][line width=0.75]  (468,3191.5) -- (498,3191.5) -- (498,3221.25) -- (468,3221.25) -- cycle ;
\draw   (498,3132) -- (528,3132) -- (528,3161.75) -- (498,3161.75) -- cycle ;
\draw   (498,3161.75) -- (528,3161.75) -- (528,3191.5) -- (498,3191.5) -- cycle ;
\draw   (498,3132) -- (528,3132) -- (528,3161.75) -- (498,3161.75) -- cycle ;
\draw   (498,3221.25) -- (528,3221.25) -- (528,3251) -- (498,3251) -- cycle ;
\draw   (498,3251) -- (528,3251) -- (528,3280.75) -- (498,3280.75) -- cycle ;
\draw   (498,3251) -- (528,3251) -- (528,3280.75) -- (498,3280.75) -- cycle ;
\draw   (498,3310.5) -- (528,3310.5) -- (528,3340.25) -- (498,3340.25) -- cycle ;
\draw   (498,3340.25) -- (528,3340.25) -- (528,3370) -- (498,3370) -- cycle ;
\draw  [color={rgb, 255:red, 65; green, 117; blue, 5 }  ,draw opacity=1 ][line width=2.25]  (138,3191.5) -- (168,3191.5) -- (168,3221.25) -- (138,3221.25) -- cycle ;
\draw  [color={rgb, 255:red, 65; green, 117; blue, 5 }  ,draw opacity=1 ][line width=2.25]  (138,3102.25) -- (168,3102.25) -- (168,3132) -- (138,3132) -- cycle ;
\draw  [color={rgb, 255:red, 65; green, 117; blue, 5 }  ,draw opacity=1 ][line width=2.25]  (138,3280.75) -- (168,3280.75) -- (168,3310.5) -- (138,3310.5) -- cycle ;
\draw  [color={rgb, 255:red, 65; green, 117; blue, 5 }  ,draw opacity=1 ][line width=2.25]  (138,3370) -- (168,3370) -- (168,3399.75) -- (138,3399.75) -- cycle ;
\draw  [color={rgb, 255:red, 65; green, 117; blue, 5 }  ,draw opacity=1 ][line width=2.25]  (258,3370) -- (288,3370) -- (288,3399.75) -- (258,3399.75) -- cycle ;
\draw  [color={rgb, 255:red, 65; green, 117; blue, 5 }  ,draw opacity=1 ][line width=2.25]  (378,3370) -- (408,3370) -- (408,3399.75) -- (378,3399.75) -- cycle ;
\draw  [color={rgb, 255:red, 65; green, 117; blue, 5 }  ,draw opacity=1 ][line width=2.25]  (498,3370) -- (528,3370) -- (528,3399.75) -- (498,3399.75) -- cycle ;
\draw  [color={rgb, 255:red, 65; green, 117; blue, 5 }  ,draw opacity=1 ][line width=2.25]  (498,3280.75) -- (528,3280.75) -- (528,3310.5) -- (498,3310.5) -- cycle ;
\draw  [color={rgb, 255:red, 65; green, 117; blue, 5 }  ,draw opacity=1 ][line width=2.25]  (378,3280.75) -- (408,3280.75) -- (408,3310.5) -- (378,3310.5) -- cycle ;
\draw  [color={rgb, 255:red, 65; green, 117; blue, 5 }  ,draw opacity=1 ][line width=2.25]  (258,3280.75) -- (288,3280.75) -- (288,3310.5) -- (258,3310.5) -- cycle ;
\draw  [color={rgb, 255:red, 65; green, 117; blue, 5 }  ,draw opacity=1 ][line width=2.25]  (258,3191.5) -- (288,3191.5) -- (288,3221.25) -- (258,3221.25) -- cycle ;
\draw  [color={rgb, 255:red, 65; green, 117; blue, 5 }  ,draw opacity=1 ][line width=2.25]  (378,3191.5) -- (408,3191.5) -- (408,3221.25) -- (378,3221.25) -- cycle ;
\draw  [color={rgb, 255:red, 65; green, 117; blue, 5 }  ,draw opacity=1 ][line width=2.25]  (498,3191.5) -- (528,3191.5) -- (528,3221.25) -- (498,3221.25) -- cycle ;
\draw  [color={rgb, 255:red, 65; green, 117; blue, 5 }  ,draw opacity=1 ][line width=2.25]  (378,3102.25) -- (408,3102.25) -- (408,3132) -- (378,3132) -- cycle ;
\draw  [color={rgb, 255:red, 65; green, 117; blue, 5 }  ,draw opacity=1 ][line width=2.25]  (498,3102.25) -- (528,3102.25) -- (528,3132) -- (498,3132) -- cycle ;
\draw  [color={rgb, 255:red, 65; green, 117; blue, 5 }  ,draw opacity=1 ][line width=2.25]  (258,3102.25) -- (288,3102.25) -- (288,3132) -- (258,3132) -- cycle ;

\draw (121,3371.5) node [anchor=north west][inner sep=0.75pt]   [align=left] {{\fontfamily{pcr}\selectfont 0}};
\draw (122,3343.5) node [anchor=north west][inner sep=0.75pt]   [align=left] {{\fontfamily{pcr}\selectfont 1}};
\draw (121,3313.5) node [anchor=north west][inner sep=0.75pt]   [align=left] {{\fontfamily{pcr}\selectfont 2}};
\draw (121,3282.5) node [anchor=north west][inner sep=0.75pt]   [align=left] {{\fontfamily{pcr}\selectfont 3}};
\draw (121,3251.5) node [anchor=north west][inner sep=0.75pt]   [align=left] {{\fontfamily{pcr}\selectfont 4}};
\draw (121,3224.5) node [anchor=north west][inner sep=0.75pt]   [align=left] {{\fontfamily{pcr}\selectfont 5}};
\draw (121,3195.5) node [anchor=north west][inner sep=0.75pt]   [align=left] {{\fontfamily{pcr}\selectfont 6}};
\draw (120,3167.5) node [anchor=north west][inner sep=0.75pt]   [align=left] {{\fontfamily{pcr}\selectfont 7}};
\draw (120,3137.5) node [anchor=north west][inner sep=0.75pt]   [align=left] {{\fontfamily{pcr}\selectfont 8}};
\draw (120,3106.5) node [anchor=north west][inner sep=0.75pt]   [align=left] {{\fontfamily{pcr}\selectfont 9}};
\draw (144,3400.5) node [anchor=north west][inner sep=0.75pt]   [align=left] {{\fontfamily{pcr}\selectfont 0}};
\draw (175,3400.5) node [anchor=north west][inner sep=0.75pt]   [align=left] {{\fontfamily{pcr}\selectfont 1}};
\draw (205,3400.5) node [anchor=north west][inner sep=0.75pt]   [align=left] {{\fontfamily{pcr}\selectfont 2}};
\draw (235,3400.5) node [anchor=north west][inner sep=0.75pt]   [align=left] {{\fontfamily{pcr}\selectfont 3}};
\draw (264,3399.75) node [anchor=north west][inner sep=0.75pt]   [align=left] {{\fontfamily{pcr}\selectfont 4}};
\draw (296,3400.5) node [anchor=north west][inner sep=0.75pt]   [align=left] {{\fontfamily{pcr}\selectfont 5}};
\draw (325,3400.5) node [anchor=north west][inner sep=0.75pt]   [align=left] {{\fontfamily{pcr}\selectfont 6}};
\draw (357,3400.5) node [anchor=north west][inner sep=0.75pt]   [align=left] {{\fontfamily{pcr}\selectfont 7}};
\draw (386,3400.5) node [anchor=north west][inner sep=0.75pt]   [align=left] {{\fontfamily{pcr}\selectfont 8}};
\draw (416,3400.5) node [anchor=north west][inner sep=0.75pt]   [align=left] {{\fontfamily{pcr}\selectfont 9}};
\draw (442,3400.5) node [anchor=north west][inner sep=0.75pt]   [align=left] {{\fontfamily{pcr}\selectfont 10}};
\draw (472,3400.5) node [anchor=north west][inner sep=0.75pt]   [align=left] {{\fontfamily{pcr}\selectfont 11}};
\draw (503,3400.5) node [anchor=north west][inner sep=0.75pt]   [align=left] {{\fontfamily{pcr}\selectfont 12}};
\draw (137,3107.5) node [anchor=north west][inner sep=0.75pt]  [font=\footnotesize] [align=left] {$\displaystyle \ket{1_{L}}$};
\draw (260,3107.5) node [anchor=north west][inner sep=0.75pt]  [font=\footnotesize] [align=left] {$\displaystyle \ket{1_{L}}$};
\draw (379,3108.5) node [anchor=north west][inner sep=0.75pt]  [font=\footnotesize] [align=left] {$\displaystyle \ket{1_{L}}$};
\draw (498,3108.5) node [anchor=north west][inner sep=0.75pt]  [font=\footnotesize] [align=left] {$\displaystyle \ket{1_{L}}$};
\draw (139,3287.5) node [anchor=north west][inner sep=0.75pt]  [font=\footnotesize] [align=left] {$\displaystyle \ket{1_{L}}$};
\draw (258,3287.5) node [anchor=north west][inner sep=0.75pt]  [font=\footnotesize] [align=left] {$\displaystyle \ket{1_{L}}$};
\draw (380,3287.5) node [anchor=north west][inner sep=0.75pt]  [font=\footnotesize] [align=left] {$\displaystyle \ket{1_{L}}$};
\draw (498,3286.75) node [anchor=north west][inner sep=0.75pt]  [font=\footnotesize] [align=left] {$\displaystyle \ket{1_{L}}$};
\draw (140,3198.5) node [anchor=north west][inner sep=0.75pt]  [font=\footnotesize] [align=left] {$\displaystyle \ket{0_{L}}$};
\draw (259,3197.5) node [anchor=north west][inner sep=0.75pt]  [font=\footnotesize] [align=left] {$\displaystyle \ket{0_{L}}$};
\draw (378,3197.5) node [anchor=north west][inner sep=0.75pt]  [font=\footnotesize] [align=left] {$\displaystyle \ket{0_{L}}$};
\draw (500,3197.5) node [anchor=north west][inner sep=0.75pt]  [font=\footnotesize] [align=left] {$\displaystyle \ket{0_{L}}$};
\draw (500,3375.5) node [anchor=north west][inner sep=0.75pt]  [font=\footnotesize] [align=left] {$\displaystyle \ket{0_{L}}$};
\draw (379,3375.5) node [anchor=north west][inner sep=0.75pt]  [font=\footnotesize] [align=left] {$\displaystyle \ket{0_{L}}$};
\draw (260,3375.5) node [anchor=north west][inner sep=0.75pt]  [font=\footnotesize] [align=left] {$\displaystyle \ket{0_{L}}$};
\draw (138,3375) node [anchor=north west][inner sep=0.75pt]  [font=\footnotesize] [align=left] {$\displaystyle \ket{0_{L}}$};

\end{tikzpicture}

}

%% file: figures/continuous_X_level.tex
\tikzset{every picture/.style={line width=0.75pt}} 

\begin{tikzpicture}[x=0.75pt,y=0.75pt,yscale=-1,xscale=1]

\draw    (333,6671) -- (333,6283) ;
\draw [shift={(333,6280)}, rotate = 90] [fill={rgb, 255:red, 0; green, 0; blue, 0 }  ][line width=0.08]  [draw opacity=0] (10.72,-5.15) -- (0,0) -- (10.72,5.15) -- (7.12,0) -- cycle    ;
\draw [shift={(333,6674)}, rotate = 270] [fill={rgb, 255:red, 0; green, 0; blue, 0 }  ][line width=0.08]  [draw opacity=0] (10.72,-5.15) -- (0,0) -- (10.72,5.15) -- (7.12,0) -- cycle    ;
\draw [color={rgb, 255:red, 208; green, 2; blue, 27 }  ,draw opacity=1 ]   (310,6331) -- (355.5,6331) ;
\draw [color={rgb, 255:red, 74; green, 144; blue, 226 }  ,draw opacity=1 ]   (311,6371) -- (356.5,6371) ;
\draw [color={rgb, 255:red, 208; green, 2; blue, 27 }  ,draw opacity=1 ]   (312,6409) -- (357.5,6409) ;
\draw [color={rgb, 255:red, 74; green, 144; blue, 226 }  ,draw opacity=1 ]   (312,6445) -- (357.5,6445) ;
\draw [color={rgb, 255:red, 208; green, 2; blue, 27 }  ,draw opacity=1 ]   (312,6481) -- (357.5,6481) ;
\draw [color={rgb, 255:red, 74; green, 144; blue, 226 }  ,draw opacity=1 ]   (312,6516) -- (357.5,6516) ;
\draw [color={rgb, 255:red, 208; green, 2; blue, 27 }  ,draw opacity=1 ]   (312,6549) -- (357.5,6549) ;
\draw [color={rgb, 255:red, 74; green, 144; blue, 226 }  ,draw opacity=1 ]   (312,6586) -- (357.5,6586) ;
\draw [color={rgb, 255:red, 208; green, 2; blue, 27 }  ,draw opacity=1 ]   (314,6622) -- (359.5,6622) ;
\draw    (306,6481) -- (306,6464) ;
\draw [shift={(306,6461)}, rotate = 90] [fill={rgb, 255:red, 0; green, 0; blue, 0 }  ][line width=0.08]  [draw opacity=0] (10.72,-5.15) -- (0,0) -- (10.72,5.15) -- (7.12,0) -- cycle    ;

\draw (365,6317) node [anchor=north west][inner sep=0.75pt]    {$4|\lambda _{v} \rangle $};
\draw (366,6356) node [anchor=north west][inner sep=0.75pt]    {$3\ |\lambda _{v} \rangle $};
\draw (366,6395) node [anchor=north west][inner sep=0.75pt]    {$2\ |\lambda _{v} \rangle $};
\draw (365,6431) node [anchor=north west][inner sep=0.75pt]    {$|\lambda _{v} \rangle $};
\draw (365,6503) node [anchor=north west][inner sep=0.75pt]    {$-\ |\lambda _{v} \rangle $};
\draw (365,6535) node [anchor=north west][inner sep=0.75pt]    {$-\ 2\ |\lambda _{v} \rangle $};
\draw (365,6576) node [anchor=north west][inner sep=0.75pt]    {$-\ 3\ |\lambda _{v} \rangle $};
\draw (366,6609) node [anchor=north west][inner sep=0.75pt]    {$-\ 4\ |\lambda _{v} \rangle $};
\draw (367,6468) node [anchor=north west][inner sep=0.75pt]    {$|0\rangle $};
\draw (275,6457) node [anchor=north west][inner sep=0.75pt]    {$\delta _{Y}$};

continuous\end{tikzpicture}

%% file: figures/continuous_variables_XY.tex
\tikzset{every picture/.style={line width=0.75pt}} 

\begin{tikzpicture}[x=0.75pt,y=0.75pt,yscale=-1,xscale=1]

\draw    (333,7181) -- (333,6793) ;
\draw [shift={(333,6790)}, rotate = 90] [fill={rgb, 255:red, 0; green, 0; blue, 0 }  ][line width=0.08]  [draw opacity=0] (10.72,-5.15) -- (0,0) -- (10.72,5.15) -- (7.12,0) -- cycle    ;
\draw [shift={(333,7184)}, rotate = 270] [fill={rgb, 255:red, 0; green, 0; blue, 0 }  ][line width=0.08]  [draw opacity=0] (10.72,-5.15) -- (0,0) -- (10.72,5.15) -- (7.12,0) -- cycle    ;
\draw [color={rgb, 255:red, 208; green, 2; blue, 27 }  ,draw opacity=1 ]   (310,6841) -- (355.5,6841) ;
\draw [color={rgb, 255:red, 74; green, 144; blue, 226 }  ,draw opacity=1 ]   (311,6881) -- (356.5,6881) ;
\draw [color={rgb, 255:red, 208; green, 2; blue, 27 }  ,draw opacity=1 ]   (312,6919) -- (357.5,6919) ;
\draw [color={rgb, 255:red, 74; green, 144; blue, 226 }  ,draw opacity=1 ]   (313,6955) -- (358.5,6955) ;
\draw [color={rgb, 255:red, 208; green, 2; blue, 27 }  ,draw opacity=1 ]   (313,6991) -- (358.5,6991) ;
\draw [color={rgb, 255:red, 74; green, 144; blue, 226 }  ,draw opacity=1 ]   (313,7026) -- (358.5,7026) ;
\draw [color={rgb, 255:red, 208; green, 2; blue, 27 }  ,draw opacity=1 ]   (313,7059) -- (358.5,7059) ;
\draw [color={rgb, 255:red, 74; green, 144; blue, 226 }  ,draw opacity=1 ]   (313,7096) -- (358.5,7096) ;
\draw [color={rgb, 255:red, 208; green, 2; blue, 27 }  ,draw opacity=1 ]   (315,7132) -- (360.5,7132) ;
\draw    (369,6877) -- (369,6860) ;
\draw [shift={(369,6857)}, rotate = 90] [fill={rgb, 255:red, 0; green, 0; blue, 0 }  ][line width=0.08]  [draw opacity=0] (10.72,-5.15) -- (0,0) -- (10.72,5.15) -- (7.12,0) -- cycle    ;
\draw    (143.25,6991.82) -- (531.25,6991.68) ;
\draw [shift={(534.25,6991.68)}, rotate = 179.98] [fill={rgb, 255:red, 0; green, 0; blue, 0 }  ][line width=0.08]  [draw opacity=0] (10.72,-5.15) -- (0,0) -- (10.72,5.15) -- (7.12,0) -- cycle    ;
\draw [shift={(140.25,6991.82)}, rotate = 359.98] [fill={rgb, 255:red, 0; green, 0; blue, 0 }  ][line width=0.08]  [draw opacity=0] (10.72,-5.15) -- (0,0) -- (10.72,5.15) -- (7.12,0) -- cycle    ;
\draw [color={rgb, 255:red, 208; green, 2; blue, 27 }  ,draw opacity=1 ]   (483.24,6968.7) -- (483.26,7014.2) ;
\draw [color={rgb, 255:red, 74; green, 144; blue, 226 }  ,draw opacity=1 ]   (443.24,6969.71) -- (443.26,7015.21) ;
\draw [color={rgb, 255:red, 208; green, 2; blue, 27 }  ,draw opacity=1 ]   (405.24,6970.73) -- (405.26,7016.23) ;
\draw [color={rgb, 255:red, 74; green, 144; blue, 226 }  ,draw opacity=1 ]   (369.24,6971.74) -- (369.26,7017.24) ;
\draw [color={rgb, 255:red, 208; green, 2; blue, 27 }  ,draw opacity=1 ]   (333.24,6971.75) -- (333.26,7017.25) ;
\draw [color={rgb, 255:red, 74; green, 144; blue, 226 }  ,draw opacity=1 ]   (298.24,6971.76) -- (298.26,7017.26) ;
\draw [color={rgb, 255:red, 208; green, 2; blue, 27 }  ,draw opacity=1 ]   (265.24,6971.78) -- (265.26,7017.28) ;
\draw [color={rgb, 255:red, 74; green, 144; blue, 226 }  ,draw opacity=1 ]   (228.24,6971.79) -- (228.26,7017.29) ;
\draw [color={rgb, 255:red, 208; green, 2; blue, 27 }  ,draw opacity=1 ]   (192.24,6973.8) -- (192.26,7019.3) ;
\draw    (490,7005) -- (510.5,7005) ;
\draw [shift={(513.5,7005)}, rotate = 180] [fill={rgb, 255:red, 0; green, 0; blue, 0 }  ][line width=0.08]  [draw opacity=0] (10.72,-5.15) -- (0,0) -- (10.72,5.15) -- (7.12,0) -- cycle    ;

\draw (361,7013) node [anchor=north west][inner sep=0.75pt]    {$-\ \lambda _{v}$};
\draw (361,7045) node [anchor=north west][inner sep=0.75pt]    {$-\ 2\ \lambda _{v}$};
\draw (361,7086) node [anchor=north west][inner sep=0.75pt]    {$-\ 3\ \lambda _{v}$};
\draw (362,7119) node [anchor=north west][inner sep=0.75pt]    {$-\ 4\ \lambda _{v}$};
\draw (338,6853) node [anchor=north west][inner sep=0.75pt]    {$\delta _{Y}$};
\draw (242,6823) node [anchor=north west][inner sep=0.75pt]    {$4\ \lambda _{v}$};
\draw (243,6862) node [anchor=north west][inner sep=0.75pt]    {$3\ \lambda _{v}$};
\draw (243,6901) node [anchor=north west][inner sep=0.75pt]    {$2\ \lambda _{v}$};
\draw (250,6937) node [anchor=north west][inner sep=0.75pt]    {$\lambda _{v}$};
\draw (180.14,7123.66) node [anchor=north west][inner sep=0.75pt]  [rotate=-269.21]  {$-\ 4\ \lambda _{h}$};
\draw (216.14,7123.66) node [anchor=north west][inner sep=0.75pt]  [rotate=-269.21]  {$-\ 3\ \lambda _{h}$};
\draw (253.14,7124.66) node [anchor=north west][inner sep=0.75pt]  [rotate=-269.21]  {$-\ 2\ \lambda _{h}$};
\draw (280.14,7112.66) node [anchor=north west][inner sep=0.75pt]  [rotate=-269.21]  {$-\ \lambda _{h}$};
\draw (357.14,6969.66) node [anchor=north west][inner sep=0.75pt]  [rotate=-269.21]  {$\lambda _{h}$};
\draw (393.14,6969.66) node [anchor=north west][inner sep=0.75pt]  [rotate=-269.21]  {$2\ \lambda _{h}$};
\draw (430.14,6970.66) node [anchor=north west][inner sep=0.75pt]  [rotate=-269.21]  {$3\ \lambda _{h}$};
\draw (471.14,6964.66) node [anchor=north west][inner sep=0.75pt]  [rotate=-269.21]  {$4\ \lambda _{h}$};
\draw (491,7005) node [anchor=north west][inner sep=0.75pt]    {$\delta _{X}$};

\end{tikzpicture}

%% file: sections/conclusion.tex
\section{Conclusion}
In this paper, we offered a guide on how to build a a self-contained introduction to GKP qubits for an audience with undergraduate CS background. Each new concept was introduced using a mixture of foundational pre-requisites which are included in most standard undergraduate CS degrees and visual-based intuitive cues. We hope facilitators will find the resource useful in building their sessions and hope this will promote the study of CVQC for non-physicists.

%% file: main.bbl
\begin{thebibliography}{10}

\bibitem{A22}
V.~V. Albert.
\newblock Bosonic coding: introduction and use cases.
\newblock In {\em Quantum Fluids of Light and Matter}, pages 79--107. IOS Press, 2025.

\bibitem{Alvargonzález01122011}
D.~A. and.
\newblock Multidisciplinarity, interdisciplinarity, transdisciplinarity, and the sciences.
\newblock {\em International Studies in the Philosophy of Science}, 25(4):387--403, 2011.

\bibitem{archer2022visual}
N.~Archer.
\newblock Visual design of quantum physics – lessons learned from nine gamified and artistic quantum physics projects, 2022.

\bibitem{Deveney:23}
E.~Deveney, E.~Demirbas, and S.~Serna.
\newblock Quantum mechanics in a quicker, more intuitive, and accessible way.
\newblock In {\em Seventeenth Conference on Education and Training in Optics and Photonics: ETOP 2023}, page 1272334. Optica Publishing Group, 2023.

\bibitem{Dimitrov:2023bjw}
E.~Dimitrov, C.~Dunne, V.~Kannan, K.~Krishnakumar, P.~L.~M. de~Rituerto, P.~S. Vieites, R.~Tiwari, and R.~A. Wolf.
\newblock {QPCC: A Quantum Programming Course for Inhomogeneous Cohorts of Professional Learners}.
\newblock In {\em {2023 International Conference on Quantum Computing and Engineering}}, 9 2023.

\bibitem{folkersvisualising}
B.~Folkers, K.~Stadermann, and A.~Brinkman.
\newblock Visualising the invisible: Reviewing the literature on demonstration material for quantum entanglement, 2024.
\newblock Unpublished manuscript, presentation, or report; Accessed [Date of Access, e.g., 9 July 2025].

\bibitem{GKP01}
D.~Gottesman, A.~Kitaev, and J.~Preskill.
\newblock Encoding a qubit in an oscillator.
\newblock {\em Phys. Rev. A}, 64:012310, Jun 2001.

\bibitem{GP21}
A.~L. Grimsmo and S.~Puri.
\newblock Quantum error correction with the gottesman-kitaev-preskill code.
\newblock {\em PRX Quantum}, 2:020101, Jun 2021.

\bibitem{martonosi2019stepsquantumcomputingcomputer}
M.~Martonosi and M.~Roetteler.
\newblock Next steps in quantum computing: Computer science's role, 2019.

\bibitem{MYHK20}
T.~Matsuura, H.~Yamasaki, and M.~Koashi.
\newblock Equivalence of approximate gottesman-kitaev-preskill codes.
\newblock {\em Phys. Rev. A}, 102:032408, Sep 2020.

\bibitem{mcgrath2005visual}
M.~B. McGrath and J.~R. Brown.
\newblock Visual learning for science and engineering.
\newblock {\em IEEE Computer Graphics and Applications}, 25(5):56--63, 2005.

\bibitem{doi:10.1177/00187208211048301}
N.~Michinov and S.~Jeanson.
\newblock Creativity in scientific research: Multidisciplinarity fosters depth of ideas among scientists in electronic “brainwriting” groups.
\newblock {\em Human Factors}, 65(7):1542--1553, 2023.
\newblock PMID: 34607488.

\bibitem{nilson2016teaching}
L.~B. Nilson.
\newblock {\em Teaching at its best: A research-based resource for college instructors}.
\newblock John Wiley \& Sons, 2016.

\bibitem{raiyn2016role}
J.~Raiyn.
\newblock The role of visual learning in improving students' high-order thinking skills.
\newblock {\em Journal of Education and Practice}, 7(24):115--121, 2016.

\bibitem{rieffel2000introduction}
E.~Rieffel and W.~Polak.
\newblock An introduction to quantum computing for non-physicists.
\newblock {\em ACM Computing Surveys (CSUR)}, 32(3):300--335, 2000.

\bibitem{seegerer2021quantum}
S.~Seegerer, T.~Michaeli, and R.~Romeike.
\newblock Quantum computing as a topic in computer science education.
\newblock In {\em Proceedings of the 16th workshop in primary and secondary computing education}, pages 1--6, 2021.

\bibitem{WBZ23}
J.~Wu, A.~J. Brady, and Q.~Zhuang.
\newblock Optimal encoding of oscillators into more oscillators.
\newblock {\em {Quantum}}, 7:1082, Aug. 2023.

\end{thebibliography}
